\documentclass[%
 aip,
 amsmath,amssymb,
preprint,%
]{revtex4-2}

\usepackage{graphicx}
\usepackage{dcolumn}
\usepackage{bm}

\usepackage[utf8]{inputenc}
\usepackage[T1]{fontenc}
\usepackage{mathptmx}
\usepackage{epstopdf, epsfig}
\graphicspath{{./fig/}}
\usepackage{color}
\usepackage{amsmath}
\usepackage{tabularx}
\usepackage{float}
\usepackage{CJKutf8}

\begin{document}

\title{Capillary-driven migration of droplets on conical fibers}
\author{Yixiao Mao
	\begin{CJK}{UTF8}{gbsn}
		(毛怡霄)
	\end{CJK}
}
\author{Chengxi Zhao
	\begin{CJK}{UTF8}{gbsn}
		(赵承熙)
	\end{CJK}
}
\email[]{zhaochengxi@ustc.edu.cn}
\author{Kai Mu
	\begin{CJK}{UTF8}{gbsn}
		(穆恺)
	\end{CJK}
}
\affiliation{Department of Modern Mechanics, University of Science and Technology of China, Hefei, Anhui 230026, PR China}
\author{Kai Li
	\begin{CJK}{UTF8}{gbsn}
		(李凯)
	\end{CJK}
}
\affiliation{ School of Engineering Science, University of Chinese Academy of Sciences, Beijing 100049, PR China}
\affiliation{ Key Laboratory of Microgravity, Institute of Mechanics, Chinese Academy of Sciences, Beijing 100190, PR China}
\author{Ting Si
	\begin{CJK}{UTF8}{gbsn}
		(司廷)
	\end{CJK}
}
\affiliation{Department of Modern Mechanics, University of Science and Technology of China, Hefei, Anhui 230026, PR China}

\date{\today}

\begin{abstract}
A droplet placed on a hydrophilic conical fiber tends to move toward the end of larger radii due to capillary action. 
Experimental investigations are performed to explore the dynamics of droplets with varying viscosities and volumes on different fibers at the microscale. Droplets are found to accelerate initially and subsequently decelerate during migration. 
A dynamic model is developed to capture dynamics of the droplet migration, addressing the limitations of previous equilibrium-based scaling laws.
Both experimental results and theoretical predictions indicate that droplets on more divergent fibers experience a longer acceleration phase. 
Additionally, gravitational effects are pronounced on fibers with small cone angles, exerting a substantial influence on droplet migration even below the capillary scale. 
Moreover, droplets move more slowly on dry fibers compared to those prewetted with the same liquid, primarily attributed to the increased friction.
The experiments reveal the formation of a residual liquid film after droplet migration on dry fibers, leading to considerable volume loss in the droplets. To encompass the intricacies of migration on dry fibers, the model is refined to incorporate a higher friction coefficient and variable droplet volumes, providing a more comprehensive depiction of the underlying physics.
\end{abstract}
\maketitle

\section{introduction}
Droplets moving on fibers are ubiquitous in both nature and industry. Nature has ingeniously utilized fiber structures for various functions. For instance, spider silk fibers form periodic spindle-knots that capture and transport dew in humid air \cite{spider}, while cacti collect droplets using their multi-structural spines \cite{catus}. 
Conversely, fibers in the legs of water striders facilitate the self-removal of condensing water, preventing a significant risk for the creature \cite{Self-removal}.
Drawing inspiration from nature, researchers have designed fibers with conical shapes, gradient microchannels, and circular grooves to achieve ultrafast, long-distance droplet transport \cite{cross-cone}. 
In the realm of additive manufacturing, precise control of droplets on fibers is also critical for effective fabrication of metallic structures, especially in challenging environments such as space \cite{Electron,liquidmetal}.

The dynamics of droplets has been found to be affected by various physical factors, each adding layers of complexity to the system.
For example, both perfectly and partially wetting droplets over the capillary scale can slide on a tilted fiber \cite{gravity-slide,gravity2}.
When the tilted fiber undergoes vertically oscillating, droplets exhibit dynamic modes such as pumping, vibrating, and swinging \cite{vibrate}. 
Recent work has also demonstrated that under evaporative conditions, the internal dynamics of droplets on fibers differ significantly from those on flat surfaces, resulting in more uniform particle deposition \cite{coffee}.
Additionally, in the presence of transverse wind, multiple droplets on horizontal fibers can move while experiencing strong repulsive interactions, driven by the asymmetric wakes generated behind them \cite{crosswind}. 
Beyond these external influences, droplet motion can also be driven by substrate asymmetry. Variations in hydrophilicity, for example, cause droplets to migrate toward regions with higher wettability \cite{theta-induce,fog-catus}. Similarly, microstructures on the substrate can alter the contact angle and surface energy, or create a liquid film that facilitates faster sliding \cite{mirco-structure,micro-stru2,micropatterned}.

While numerous mechanisms contribute to droplet migration on a fiber, capillary forces arising from curvature gradients are often the dominant drivers, especially for microscale droplets. A classic example of this phenomenon is observed on conical fibers. 
The earliest investigations for a single droplet on a fiber date back nearly five decades, when Carroll found that droplets on thin cylindrical fibers adopt an equilibrium conformation, fully covering the fiber axisymmetrically, known as the barrel droplet \cite{Carroll-rollup}. Carroll further developed a theoretical framework for barrel droplets, assuming constant surface curvature \cite{carroll-barrel}. 
Subsequent investigations by Lorenceau and Quéré examined droplet movement on a copper cone with a fixed angle at submillimeter scales, revealing that droplets migrate toward regions of larger radii at decreasing velocities \cite{lq-wire}. They identified the gradient of Laplace pressure, or capillary force, as the primary driving force, with resistance mainly attributed to global viscous dissipation, leading to a velocity scaling law based on the balance of these forces.
Li and Thoroddsen further refined this understanding through experiments on microscale glass fibers, extending the scaling laws across different dissipation regimes \cite{li-fiber}.
Recent studies have also established additional scaling laws derived from extensive experimental investigations\cite{law_f,law_v}.
Meanwhile modeling of droplet profiles on conical fibers has been advanced through detailed analyses of surface energy \cite{surface2-tube,surface-ene3}.
However, conventional models often approximate droplet shapes and overlook intricate deformations during migration, which may limit the precision of scaling laws for long-distance droplet movement.
To address these limitations, Chan $et~al.$ developed a lubrication model to elucidate the prolonged evolution of droplet migration on conical fibers with matched asymptotic expansions \cite{chan-drop}. Their model highlights the mismatch between apparent and equilibrium contact angles, which generates a large pressure gradient at the contact line region. 
They also demonstrated that the thickness of deposited films on prewetted conical fibers varies significantly with fiber radius, droplet size and silp length \cite{chan-coating-prf,chan-film}.

Notably, prior studies have primarily focused on fibers covered by prewet films, often simplifying them as cones with fixed angles. However, real-world fibers frequently encounter dry conditions and exhibit curvature gradients, which profoundly influence the dynamics of droplet migration. Additionally, though earlier experimental findings have validated various scaling laws \cite{lq-wire,li-fiber,law_f,law_v}, they have not been systematically compared with theoretical models to assess the long-term evolution of droplet migration. These unresolved issues collectively motivate the current study.

This study presents a combined experimental and theoretical investigation into the migration of barrel-shaped droplets on conical fibers with varying curved profiles, examining both wet and dry fiber surfaces. A dynamic model is developed to characterize long-distance movements and elucidate the underlying physics governing this migration.
In Sec.\,\ref{sec:exp}, we present the experimental settings including fiber production, droplet generation and data capturing. The theoretical model consisting of capillary and viscous forces is introduced in Sec.\,\ref{sec.model}. 
Both experimental and theoretical results are displayed in Sec.\,\ref{sec.result}, while the dynamic model is firstly verified in Sec.\,\ref{sec.verf}. The effects of fiber shapes and gravity are explored in Sec.\,\ref{sec.cone} and Sec.\,\ref{sec.g} respectively. The influence of dry/wet fibers is discussed in Sec.\,\ref{sec.dry}.

\section{\label{sec:exp}Experiment setup}

The conical microfibers are produced by heating and pulling glass capillaries (Borosilicate Glass, 1.5\,mm outer diameter, 0.86\,mm inner diameter and 10\,cm length, Sutter Instrument) using a micropipette puller (P-1000 Pipette Puller, Sutter Instrument). The capillary tube is divided into two symmetrical parts with conical tips, and broken at the desired position by the edge of a coverslip on a microscope stage. The puller settings ensure precise control over the shapes of the conical tubes. In our experiments, the typical conical fiber has a diverging shape with a starting radius $\sim10\,\rm{\mu m}$ and expanding to $\sim100\,\rm{\mu m}$ over a length of $\sim5\,\rm{mm}$, and an increasing cone semi-angle from $\sim0.005\,\rm{rad}$ to $\sim0.05\,\rm{rad}$. The pulled fibers have a roughness of a few nanometers due to heating-induced polishing, which is considered smooth for droplet motion \cite{li-fiber}. The surface can be wetted by droplets of silicone oil (Dow Corning, $10$ - $1000$\,cSt).

\begin{figure}
	\includegraphics[width=0.8\linewidth]{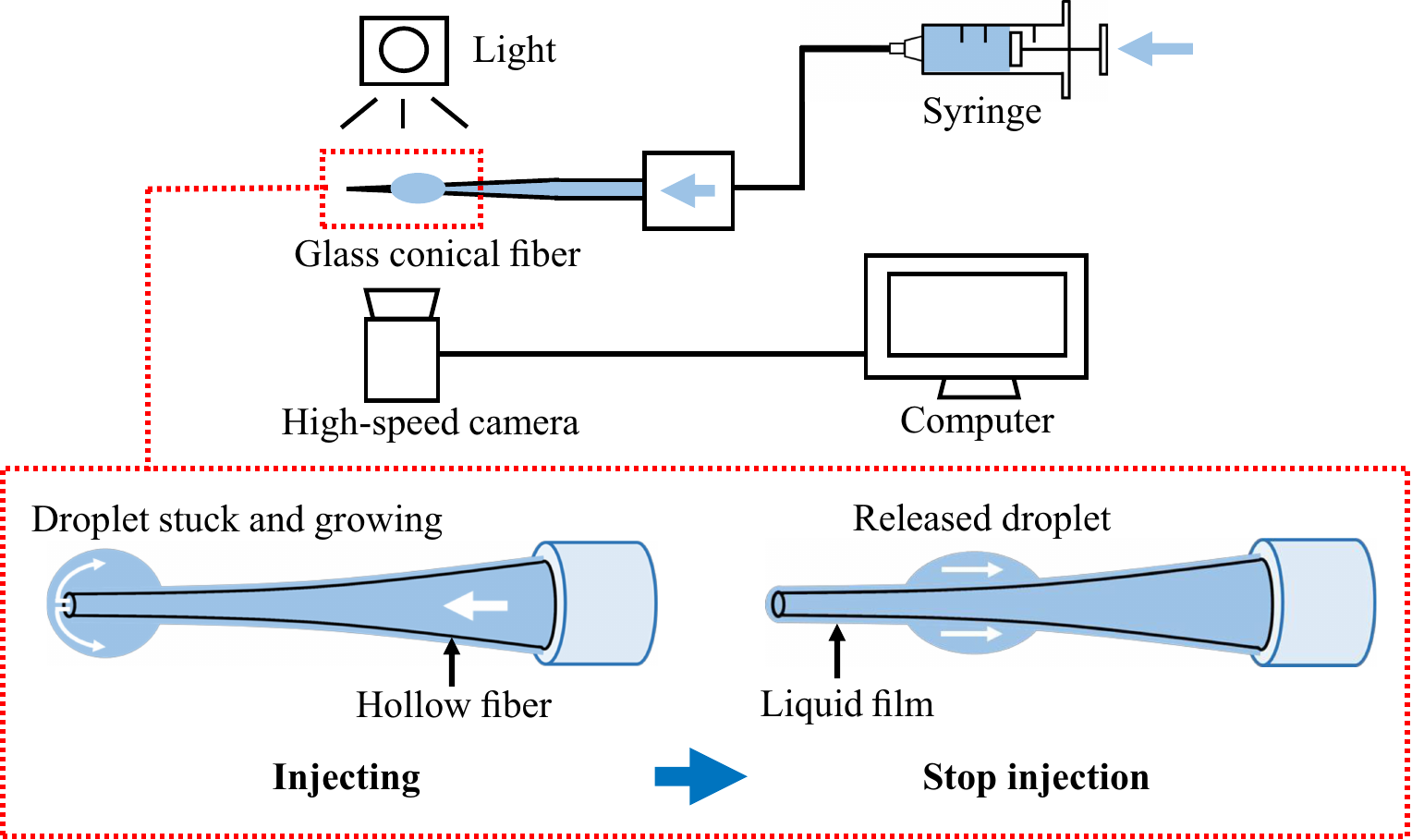}
	\caption{\label{fig:generate} Sketch of the experimental system. The detailed droplet generation on a conical fiber is highlighted within the red dotted box. Liquid flows through the catheter into a larger end of the hollow fiber under pressure from a connected syringe, accumulating at the tip of the fiber. When the pressure is released, the droplet detaches from the tip and migrates towards the thicker end of the fiber due to the curvature gradient.}
\end{figure}

Figure\,\ref{fig:generate} shows the droplet generation process. 
When the syringe is pushed, liquid is injected into a glass tube through a catheter and exits through the opening tip, forming a growing droplet attached to the outer glass wall. As continuous injection pressure is applied, the droplet remains at the tip, increasing in volume. Once the pressure is halted, the droplet detaches from the tip due to the influence of curvature gradient and begins to migrate.
This process occurs over several seconds, producing droplets with volumes ranging from $0.1$ to $100\,\rm{nL}$. When a droplet migrates along a dry fiber, it deposits a thin liquid film several micrometers thick at the receding contact line. After one or two droplet movements, the fiber becomes prewetted with a stable film, which maintains a nearly constant droplet volume during subsequent migrations. Detailed investigations are presented in Sec.\,\ref{sec.dry}.

Images of droplet migration are captured by a high-speed camera (Photron Nova S16) at rates up to 1000\,fps and a resolution of $256\times1024$ pixels, with backlighting through a diffuser. The profiles of the droplet-fiber system during migration are measured using a brightness threshold and edge capturing function in MATLAB, with an accuracy of $\sim \pm 1\,\rm{\mu m}$. 
Geometrical parameters, such as droplet volume $\Omega$ and centriod position $z_c$, are determined by comparing video frames with initial reference images captured prior to droplet deposition.

\section{\label{sec.model}Dynamic model}
\begin{figure}
    \includegraphics[width=0.8\linewidth]{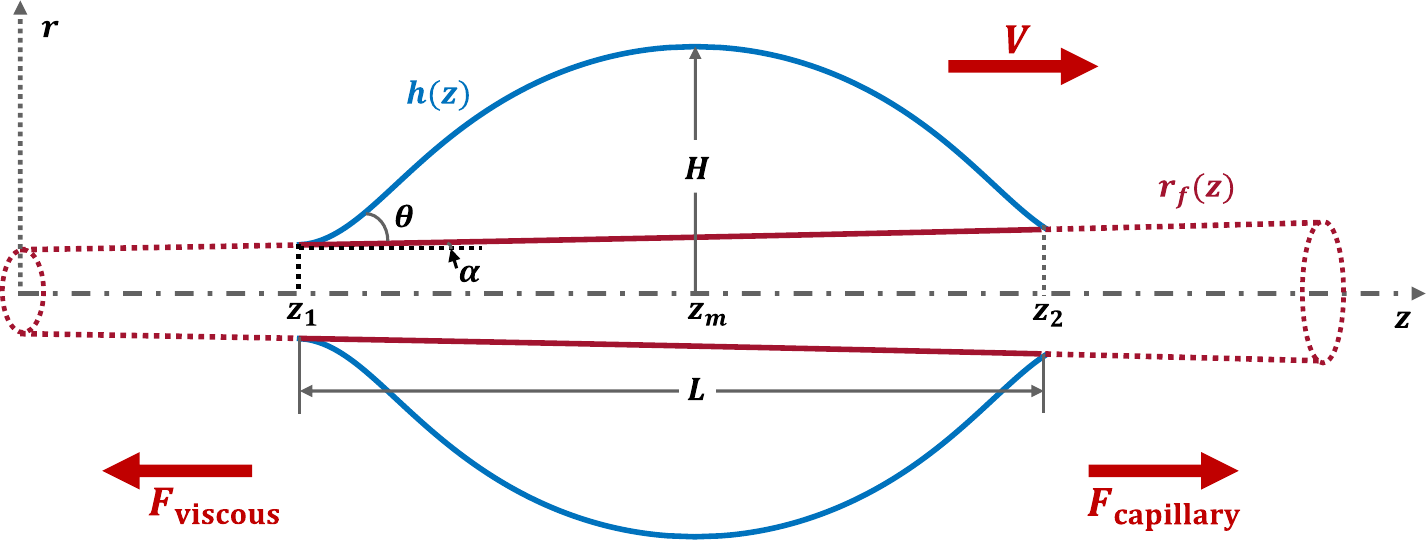}
    \caption{\label{fig:parameter} Schematic of droplet migration on a conical fiber.}
\end{figure}

Consider an axisymmetric droplet on a smooth conical fiber (Fig.\,\ref{fig:parameter}). For small cone angles, the droplet assumes a barrel shape over extended distances. The driving forces arise primarily from capillary forces due to variations in fiber radius and angle, while viscous dissipation providing resistance. 
The droplet profile is denoted by $h(z)$, with $[z_1,r_f(z_1)]$ and $[z_2,r_f(z_2)]$ representing the receding and advancing contact lines respectively. 
Additional parameters describing the droplet-fiber system include $H$ for the maximal droplet thickness at position $z_m$, $L$ for the droplet length, $\alpha=dr_f/dz$ for the fiber semi-angle, and $\theta$ for the receding contact angle. The characteristic length of the droplet is defined as $R_0=\root 3 \of {3\Omega/4\pi}$. Note that droplets are not perfectly spherical, so $R_0$ represents ``equivalent" radius. In our experiments $R_0$ ranges from 30 to 300 ${\rm\mu m}$.
All physical quantities are expressed in the International System of Units, with length measured in micrometers, time in seconds, and mass in kilograms.
Given the low capillary number $Ca\sim O(10^{-3})$ and Reynolds number $Re\sim O(10^{-4}$ - $10^{-2})$, internal flow effects are negligible. Thus the model of droplet motion is governed by Newton's law:
\begin{equation}
    \rho\Omega\frac{dV}{dt}=F_c-F_v\pm \rho g\Omega,
    \label{eq.newtonstart}
\end{equation}
where $\rho$ is the density of the liquid, $V$ is the axial velocity of the droplet centroid, $F_c$ is the capillary force, $F_v$ is the viscous force and $g$ is the gravitational acceleration, which depends on the fiber orientation.

\subsection{Capillary force}
The capillary force $F_c$ is derived from the gradient of the Laplace pressure $p$ within the droplet at different positions along the fiber, acting as the driving force, expressed as
\begin{equation}
    F_c=\frac{d p}{dz_c}\Omega,
    \label{eq.cstart}
\end{equation}
where $z_c$ is the axial position of the droplet centroid. For a droplet at an arbitrary position on the fiber, $p$ can be modeled by the Laplace equation\cite{capillary-wetting}
\begin{equation}
    p=\sigma\left[\frac{1}{h(1+h'^2)^{1/2}} - \frac{h''}{(1+h'^2)^{3/2}}\right]={\rm constant},
    \label{eq.d2}
\end{equation}
where $\sigma$ is the liquid surface tension and the prime denotes differentiation with respect to $z$.
This yields an ordinary differential equation for $h$ with respect to $z$.
Based on the schematic in Fig.\,\ref{fig:parameter}, the boundary conditions are $h=r_1, h'=\tan(\alpha_1+\theta)$ at $z=z_1$ and $h=H, h'=0$ at $z=z_m$. Here, $\alpha_1$ and $r_1$ are the fiber angle and radius at $z=z_1$ respectively.
Additionally, we assume $\theta=0$ since the receding contact line is always covered by a thin film in our experiment. After substituting the boundary conditions, Eq.\,(\ref{eq.d2}) simplifies to
\begin{equation}
	\label{eq_dh_dz}
    \frac{dh}{dz}=\pm\frac{\sqrt{(H^2-h^2)(h^2-\beta^2r_1^2)}}{h^2+\beta r_1H}~,\,{\rm where}\,\,\beta=\frac{H\cos(\alpha_1+\theta)-r_1}{H-r_1\cos(\alpha_1+\theta)}.
\end{equation}
The sign is positive before the highest point $z=z_m$ and negative subsequently. 
Solving Eq.\,\eqref{eq_dh_dz} yields the final implicit solution \cite{abramowitz1948handbook}, expressed as
\begin{equation}
 	\beta r_1F(\varphi,k)+HE(\varphi,k)-|z-z_m|=0\,,
\end{equation}
where
$$\varphi=\arcsin\sqrt{\frac{H^2-h^2}{H^2-\beta^2r_1^2}}~,~~ k=\sqrt{\frac{H^2-\beta^2r_1^2}{H^2}}\,.$$
Here $F(\varphi, k)$ and $E(\varphi, k)$ are the first and second elliptic integrals respectively,
\begin{equation}
	F(\varphi, k)=\int^{\varphi}_{0}d\phi/\sqrt{1-k^2\sin^2\phi},\quad
	E(\varphi, k)=\int^{\varphi}_{0}\sqrt{1-k^2\sin^2\phi}\,d\phi \,.
\end{equation}
Additionally, $H$ is determined by the initial droplet volume $\Omega$.
Once the fiber shape $r_f(z)$, the droplet volume $\Omega$, and the left edge $z_1$ are specified, the droplet profile $h(z)$ and the Laplace pressure $p$ are determined.
The centroid $z_c$ of the droplet is then determined by
\begin{equation}
    z_c=\int_{z_1}^{z_2}2\pi(h^2-r_f^2)z dz/\Omega\,.
\end{equation}
Calculating $p$ at each $z_c$ gives us the capillary force $F_c$.

\subsection{\label{sec.fv}Viscous force}
The resistance $F_v$ to the droplet motion arises primarily from viscous drag near the contact line at a mesoscopic scale, calculated as
\begin{eqnarray}
    F_v=\int_{z_1}^{z_2}{2\pi r_f\tau dz},
    \label{eq.fv1}
\end{eqnarray}
where $\tau$ is the shear stress at the contact area between the droplet and the fiber. According to Huh and Scriven's hydrodynamic model \cite{huh-creep}, for two dimensional steady flow in a liquid wedge with an incline angle $\theta_w$ and velocity $V$, the shear stress is approximated as $\tau=3\mu V/z_w\theta_w$. Here $z_w$ is the distance to the three-phase contact line, and $\mu$ is the liquid viscosity. The droplet on a fiber is considered as two axisymmetric wedges, each with a length of $L/2$ and angle $\theta_w\approx 2(H-r_f)/L$. Therefore, the viscous resistance $F_v$ according to Eq.\,(\ref{eq.fv1}) is expressed as
\begin{eqnarray}
    F_v=2\int_{l_{\rm min}}^{L/2}{2\pi r_f\frac{3\mu V}{z_w\theta_w}dz_w}=C_v\mu r_f\frac{L}{H-r_f}V,
    \label{eq.fv}
\end{eqnarray}
where $l_{\rm min}$ is a cutoff length introduced to aviod the logarithmic divergence\cite{lmin}, and $C_v=6\pi\ln(L/2l_{\rm min})$ is a dimensionless friction coefficient. 
Since the resistance can vary depending on the specific liquid and surface combination in the case of dynamic wetting \cite{friction}, the coefficient is usually determined via experiments. 

After calculating $F_c$ and $F_v$, Eq.\,(\ref{eq.newtonstart}) can be solved (with initial position $z_0$ and velocity $V_0$) using the solver ode45 in MATLAB to provide the dynamics of droplet migration.

\section{\label{sec.result}results and discussion}
In this section, the experimental results are presented to explore the dynamics of droplets with different viscosities and volumes on various fibers. Additionally, we utilize our dynamic model to predict droplet migration on conical fibers, demonstrating how different physical factors influence droplet behavior.
We begin by validating the dynamic model against experimental data using silicone oil droplets of differing velocities in Sec.\,\ref{sec.verf}. Subsequently, we explore the impacts of fiber shapes in Sec.\,\ref{sec.cone} and gravitational effects in Sec.\,\ref{sec.g}. The condition of fiber wetness is discussed in Sec.\,\ref{sec.dry}.

\subsection{\label{sec.verf}Model verification}
\begin{figure}[h]
	\includegraphics[width=1\linewidth]{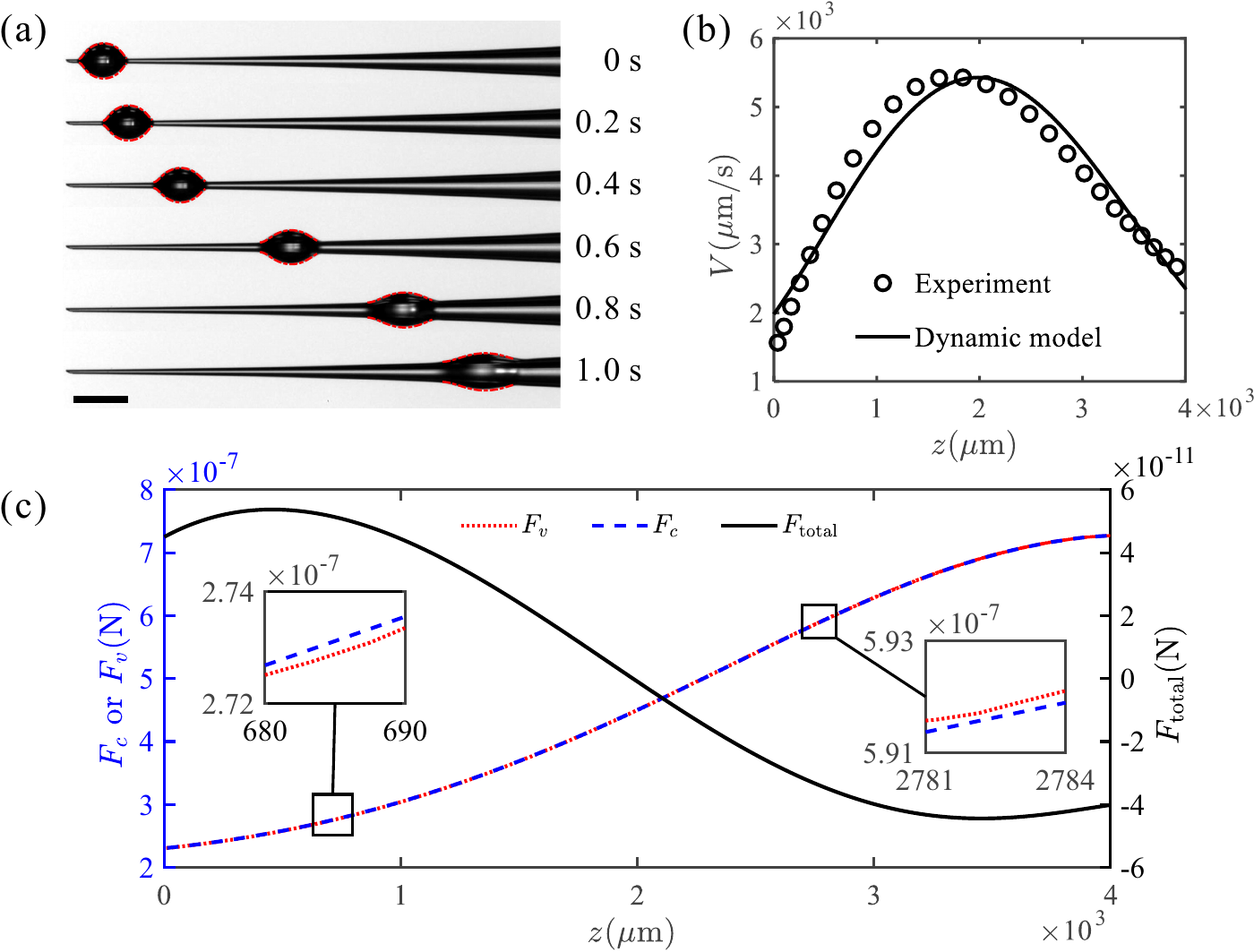}
	\caption{\label{fig:force} (a) Images of a 10 cSt silicone oil droplet migrating on a horizontal conical fiber. Red dashed lines represent the droplet shapes predicted by the theoretical model. (b) Velocity of the droplet obtained from experiment (hollow circles) and dynamic model (solid line). (c) Capillary force $F_{c}$ (blue dashed line), viscous force $F_{v}$ (red dotted line) and total force $F_{\rm total}$ (black solid line) acting on the droplet during migration. Insets highlight detailed distinctions. The initial droplet radius $R_0 = 175 ~\rm{\mu m}$. Scale bar: $500~\rm{\mu m}$.}
\end{figure}

Experimental snapshots during droplet migration on a prewetted horizontal fiber are shown in Fig.\,\ref{fig:force}(a). The droplet moves towards the thicker end, transitioning from a near-spherical to a flattened shape. 

Quantitative analysis of droplet velocity evolution $V(z)$ is presented in Fig.\,\ref{fig:force}(b), revealing distinct phases of acceleration and deceleration.
By fitting $C_v=25$ using Eq.\,(\ref{eq.newtonstart}), theoretical shapes and velocity evolution derived by the model closely match experimental observations, as depicted in Fig.\,\ref{fig:force}(a,b). 
Additionally, Fig.\,\ref{fig:force}(c) compares the capillary force $F_c$ and viscous drag $F_v$ predicted by the dynamic model.
Despite both forces increasing nearly to equilibrium during migration, the slight deviation $F_{\rm total}=F_c-F_v$ leads to significant variations in velocity.

\begin{figure}
    \includegraphics[width=0.9\linewidth]{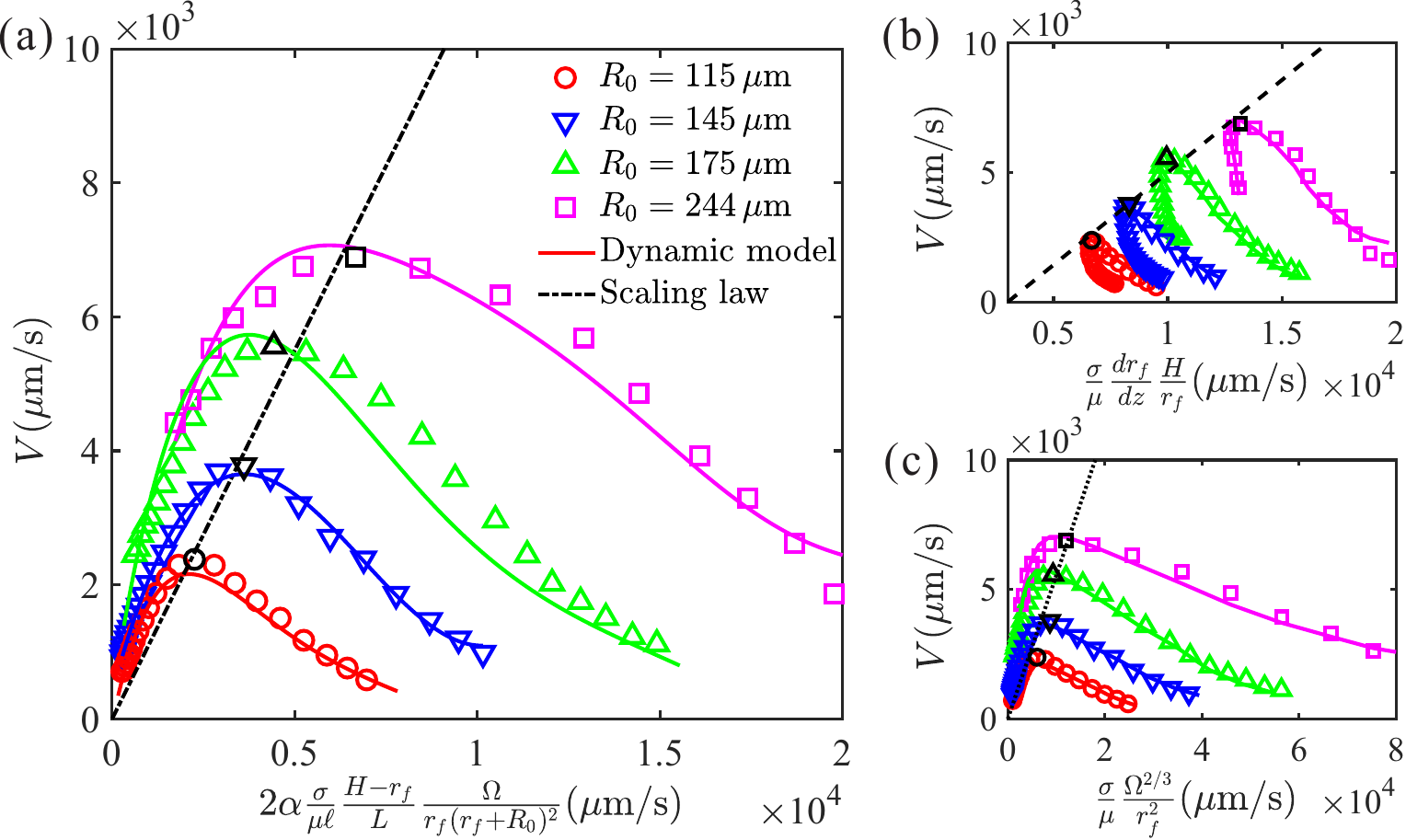}
	\caption{\label{fig:scalinglaw} Comparison of velocities from experimental results (colored symbols), the dynamic model (colored curves) and the scaling laws (black straight lines) for various droplet sizes. The black symbols represent the maximum values of velocity. The coordinates are scaled according to the laws from (a) Li and Thoroddsen \cite{li-fiber}, (b) Fournier $et\,\,al.$\cite{law_f}, and (c) Van Hulle $et\,\,al.$\cite{law_v}.} 
\end{figure}
    
We also compare our model with the scaling laws proposed by Li and Thoroddsen \cite{li-fiber}, Fournier \textit{et al}. \cite{law_f}, and Van Hulle \textit{et al}. \cite{law_v}, which balance capillary and viscous forces while neglecting gravitational effects. For the parameters in our experiment, these laws are expressed as follows:

$$V\sim\frac{\alpha\sigma}{\mu} \frac{H-r_f}{L} \frac{\Omega}{r_f(r_f+R_0)^2}~, \quad
V\sim\frac{\sigma}{\mu} \frac{dr_f}{dz} \frac{H}{r_f}~\quad \textrm{and} \quad
V\sim\frac{\sigma}{\mu} \frac{\Omega^{2/3}}{r_f^2}~.$$
Figure\,\ref{fig:scalinglaw} illustrates that these scaling laws align only with the maximum point of experimental results, while our dynamic model effectively captures the extended migration dynamics of droplets. This highlights the model's ability to surpass limitations inherent in the equilibrium assumption of scaling laws. Additionally, significant differences are observed with different droplet sizes $R_0$ (see more discussion in Appendix\,\ref{app.size}).

Since complicated factors such as the slip interface, thermal activation and surface adaptation are involved in viscous dissipation, determining the friction coefficient $C_v$ is crucial for further investigations.
We conduct experiments using silicone oil with different viscosities, and the corresponding fitted values of $C_v$ are presented in Table \ref{tab:table1}. Intriguingly, $C_v$ appears to remain unaffected by the liquid viscosity, exhibiting a consistent value ($C_v\approx25$) for droplets of silicone oil on prewetted glass fibers.

\begin{table}
  \caption{\label{tab:table1}Properties of the different liquids used in the experiments with prewetted fibers. Room temperature was kept at $23\rm{^{\circ} C}$.}
  \begin{ruledtabular}
  \begin{tabular}{ccccc}
  Liquid&$\sigma\rm{(mN/m)}$&$\mu\rm{(mPa\cdot\rm s)}$&$\rho\rm{(kg/m^3)}$&$C_v$\\
  \hline
  10 cSt	silicone oil&19.77&9.88&934&$23\pm4$\\
  50 cSt	silicone oil&20.04&48&960&$19\pm5$\\
  100 cSt	silicone oil&20.53&85&966&$26\pm10$\\
  200 cSt	silicone oil&20.64&180&980&$21\pm1$\\
  1000 cSt silicone oil&21.00&970	&980&$22\pm4$\\
  \end{tabular}
  \end{ruledtabular}
\end{table}

Our model is primarily designed to predict the dynamic behavior of small droplets during migration. When $R_0 \gg 1$\,mm, achieving uniform droplet coverage over the fiber becomes challenging, and significant inertial effects may induce interface oscillations that the model does not account for. Additionally, the model is only applicable for small contact angles, where droplets are assumed to ``slide'' along the fiber. It does not address the potential ``rolling'' behavior that may occur at larger contact angles.

\subsection{\label{sec.cone}Effects of conical shape}
\begin{figure}
    \includegraphics[width=1\linewidth]{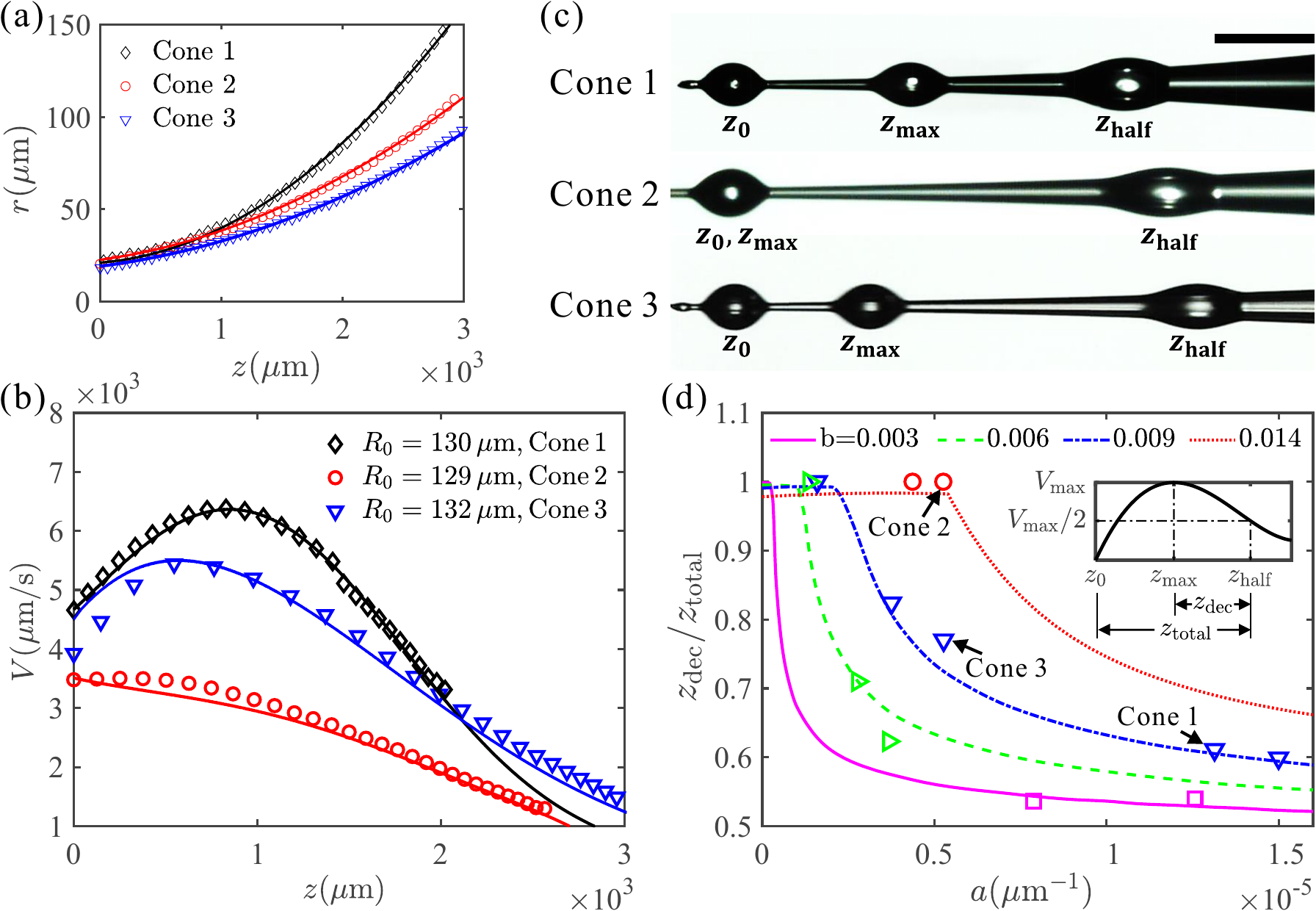} 
    \caption{\label{fig:cone} (a) Three typical fiber shapes measured in the experiments (hollow symbols). The solid lines are fitted curves of $r_f = az^2+bz+c$. (b) Velocity of droplets with similar volumes on different cones. Symbols for experiments and solid lines for dynamic model. (c) Superpositions of cones with droplets at the positions of the start $z_0$, the highest velocity $z_{\rm max}$ and half of the highest velocity $z_{\rm half}$. Scale bar is $500\,\rm{\mu m}$. (d) The ratio $z_{\rm dec}/z_{\rm total}$ as a function of the cone angle growth rate $a$. Lines are calculated by the dynamic model for the initial cone angle $b=0.003,\,0.006,\,0.009,\,0.014$ and initial radius $c=20~\rm{\mu m}$. Symbols represent the experiment data. The inset shows parameters of a typical migration curve, defining the deceleration length $z_{\rm dec}=z_{\rm half}-z_{\rm max}$ and total length $z_{\rm total}=z_{\rm half}-z_0$. }
\end{figure}

In this section, we examine the impact of the conical shape on droplet dynamics. Lorenceau and Qu\'er\'e \cite{lq-wire} observed that droplets always decelerated on the fiber with a constant cone angle. However, in this study, most droplets accelerate initially and then decelerate after reaching the maximum velocity on fibers with diverging shapes. To quantitatively assess the effects of fiber shapes, the fiber radius $r_f(z)$ is modelled as a parabola $r_f=az^2+bz+c$, where $a$ represents half the growth rate of the cone angle, $b$ represents the initial cone angle at $z=0$, and $c$ represents the initial radius. The study presents three typical conical fibers:
Cone 1 ($a=13.1\times 10^{-6}~\rm{\mu m}^{-1}, b=0.009, c=22~\rm{\mu m}$); Cone 2 ($a=5.26\times 10^{-6}~\rm{\mu m}^{-1}, b=0.014, c=23~\rm{\mu m}$); and Cone 3 ($a=5.27\times 10^{-6}~\rm{\mu m}^{-1}, b=0.009, c=19~\rm{\mu m}$). 
Figure\,\ref{fig:cone}(a) demonstrates that these parabolas accurately match the fiber shapes.

Figure\,\ref{fig:cone}(b) illustrates the velocity evolution of droplets with sizes $R_0\approx 130\,\rm{\mu m}$ on the three fibers. 
On Cone 1, characterized by the largest angle growth rate $a$, the droplet reaches its maximum velocity $V_{\rm max}$ before rapidly decelerating. On Cone 3 with a smaller angle growth rate, the droplet shows a shorter acceleration phase and more gradual deceleration. In contrast, no obvious acceleration process is observed on Cone 2 with a larger initial angle $b$.
Theoretically, there is always an acceleration phase (where speed increases from zero to its maximum) before any deceleration occurs. 	
However, determining the exact release moment is challenging due to the droplet's abrupt detachment, resulting in an initial velocity at $z=0$. Under certain conditions, the acceleration of the droplet can be so intense that the droplet quickly reaches its maximum speed.
Moreover, though there are minor variations in droplet sizes $R_0$ due to experimental uncertainties, the corresponding errors are considered negligible, as detailed in Appendix \ref{app.error}. 

For further investigation, several key points on the migration curve are defined, as shown in the inset of Fig.\,\ref{fig:cone}(d): $z_0$ for the initial position, $z_{\text{max}}$ where maximum velocity $V_{\text{max}}$ occurs, and $z_{\text{half}}$ where velocity decreases to $V_{\text{max}}/2$. These points are illustrated in Fig.\,\ref{fig:cone}(c). A dimensionless ratio $z_{\text{dec}}/z_{\text{total}}=(z_{\text{half}}-z_{\text{max}})/(z_{\text{half}}-z_0)$ is introduced to quantify deceleration to the total distance.
Figure\,\ref{fig:cone}(d) depicts the ratio $z_{\text{dec}}/z_{\text{total}}$ with the growth rate $a$ of cone angle, showing good agreement between experimental results (symbols) and theoretical predictions (lines).
For further comparisons, we supplement the experimental data with results from various conical fibers. Different colored symbols represent different initial cone angles: $b=0.003$ (red circles), 0.006 (blue lower triangles), 0.009 (green right triangles) and 0.014 (magenta squares), with all fibers having an initial radius $c=20~\rm{\mu m}$.
It is observed that acceleration becomes more pronounced as $a$ increases, primarily because a larger $a$ corresponds to a faster growth of $F_c$ in the early stage, leading to a prolonged increase in velocity. Subsequently, droplets experience rapid flattening on more diverging fibers, which intensifies viscous forces and results in more rapid deceleration.
Additionally, when $a$ is small, $z_{\text{dec}}/z_{\text{total}}\sim 1$, indicating minimal acceleration. This scenario is consistent with studies on fibers with a constant cone angle\cite{lq-wire,chan-drop} ($a=0$).
Meanwhile, deceleration becomes more dominant with increasing $b$. An increase in $b$ enhances the initial velocity near the start, with less deformation compared to a larger $a$, causing droplets to attain maximum speed closer to the cone tip.

\subsection{\label{sec.g}Effects of gravity}
Changes in fiber orientation are common in real droplet migration scenarios. For droplets ranging from microns to submillimeters, gravitational effects are typically ignored due to the small characteristic length $R_0$ compared to the capillary length $L_c=\sqrt{\sigma/\rho g}$. For instance, $L_c\approx 1.5 \rm{mm}$ for silicone oil. However, $F_c$ in this system mainly depends on the fiber structure as described in Eq.\,(\ref{eq.cstart}-\ref{eq_dh_dz}), denoting a magnitude of $\alpha\sigma\Omega/R_0^2$. Therefore, the ratio between gravity and $F_c$ is approximatively $R_0^2/\alpha L_c^2$.
Since our experiments are conducted on fibers with small angles $(\alpha\sim 0.01)$, gravitational effects cannot be neglected at submillimeter scales.

\begin{figure}
\includegraphics[width=1\linewidth]{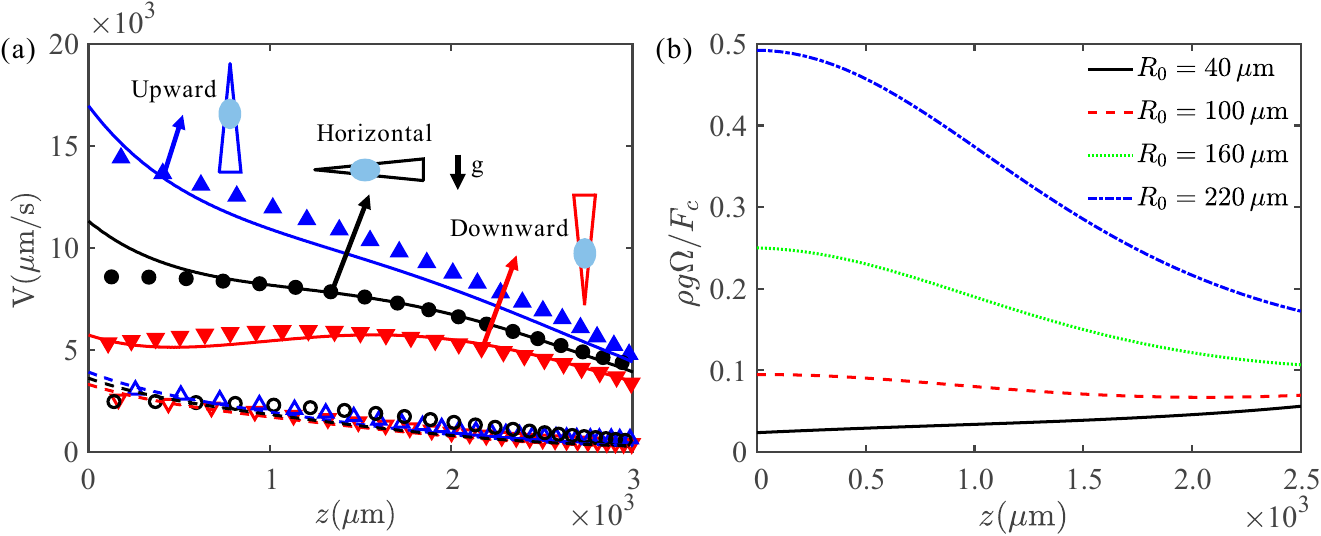}
\caption{\label{fig:gravity} (a) Velocity of 10 cSt silicone oil droplets along conical fibers positioned upwards, horizontally and downwards. Solid symbols and solid lines represent data of droplet radius $R_0 = 220~\rm{\mu m}$, while hollow symbols and dashed lines for droplet radius $R_0 = 100~\rm{\mu m}$. (b) The ratio $\rho g\Omega/F_c$ of gravitational force acting on droplets to capillary forces varies along $z$ for various droplet radii $R_0$.}
\end{figure}

In order to regulate the gravitational effects, we position the tip of the same conical fiber horizontally, vertically upwards, and vertically downwards, examining how droplet sizes influence migration.
Results of two groups of droplets with size $R_0=100~\rm{\mu m}$ (lower hollow symbols) and $R_0=220~\rm{\mu m}$ (upper solid symbols) are presented in Fig.\,\ref{fig:gravity}(a). The fiber parameters $a=4.5\times 10^{-6}~\rm{\mu m}^{-1}, b=0.014, c=26~\rm{\mu m}$, lead to continuous deceleration in horizontal orientations. 
While migration evolutions for smaller droplets ($R_0=100~\rm{\mu m}$) show minimal variation, significant differences are observed for larger droplets ($R_0=220~\rm{\mu m}$), especially near the cone tip. 
Moreover, upward velocities exceed horizontal velocities, whereas downward velocities are the lowest. This disparity arises from gravity accelerating droplets on upward fibers and decelerating them on downward fibers. 
Further quantitative assessments are conducted as described in Eq.\,(\ref{eq.newtonstart}), where the gravitational term depends on the orientation (+ for upward, - for downward). The theoretical curves plotted in Fig.\,\ref{fig:gravity}(a) closely match experimental data.
Additionally, both monotonic and non-monotonic variations of $V$ are observed depending on the orientation.
As illustrated by the red solid line in Fig.\,\ref{fig:gravity}(a), non-monotonic changes occur when the fiber is oriented downward.
In this configuration, gravity acts as a decelerating force, which suppresses initial acceleration and prolongs the acceleration phase.
Note that different gravitational orientations have minimal impact on droplet morphology (see Appendix\,\ref{app.g} for detailed discussions), further supporting the predictive capability of our model for droplet migration on fibers oriented in various directions.

Since gravity can either augment or impede the driving force, we calculate the ratio $\rho g\Omega/F_c$ during migration, as illustrated in Fig.\,\ref{fig:gravity}(b), across different droplet sizes $R_0$ and positions $z$. The ratio increases with larger droplet sizes, reflecting the escalating influence of gravity in either promoting or resisting motion. 
Furthermore, the ratio decreases from the fiber tip towards the base for $R_0 \ge 100~\rm{\mu m}$. For $R_0 = 220~\rm{\mu m}$, the maximum ratio reaches 0.5 near the fiber tip. This indicates significant disparities between upward and downward velocities, with gravity contributing to a difference of approximately 100\% of the horizontal velocity.

\subsection{\label{sec.dry}Droplet migration on a dry fiber}
\begin{figure}
\includegraphics[width=1\linewidth]{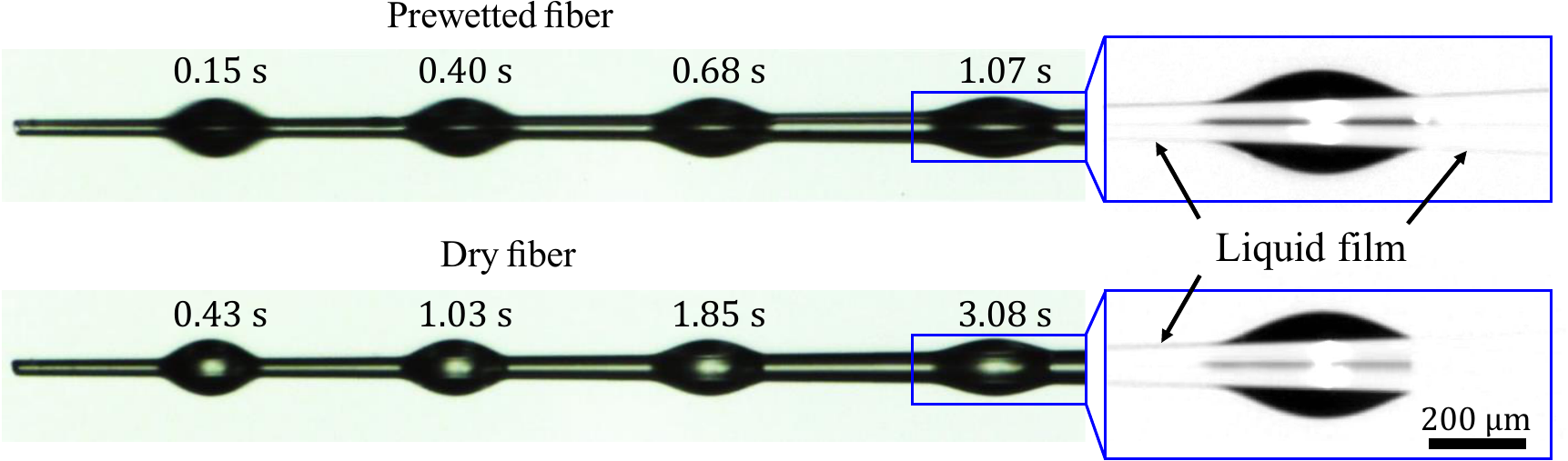}
\caption{\label{fig:film} Images of 10 cSt silicone oil droplets on a horizontal conical fiber with prewetted or dry surface. The images in the blue boxes are treated by removing the image of the dry fiber without a droplet, highlighting the geometry differences in the droplet and contact line. The liquid film is continuous along the prewetted fiber, while only exists at the receding contact line on the dry fiber.}
\end{figure}

In addition to the migration of droplets on prewetted surfaces covered by liquid films, the behavior of droplets in direct contact with dry fibers is also investigated. Figure\,\ref{fig:film} compares snapshots of the droplet migration on prewetted and dry fibers with the same geometries. Droplets on dry fibers are found to move significantly slower than those on wet fibers. Images in the blue box highlight differences in droplet and contact line between the two conditions, where the image of the dry fiber (without droplets) is excluded to emphasize the remaining liquid parts. On the prewetted fiber, a continuous liquid film is visible at both advancing and receding contact lines, while on the dry fiber the liquid film is present only at the receding contact line, a consequence of the droplet leaving liquid behind as it moves. 
Since the fiber surface is hydrophilic, the macroscopic contact angle of the droplet on the fiber is relatively small. As a result, there is no significant difference in the droplet morphology between prewetted and dry fibers.

\begin{figure}
\includegraphics[width=1\linewidth]{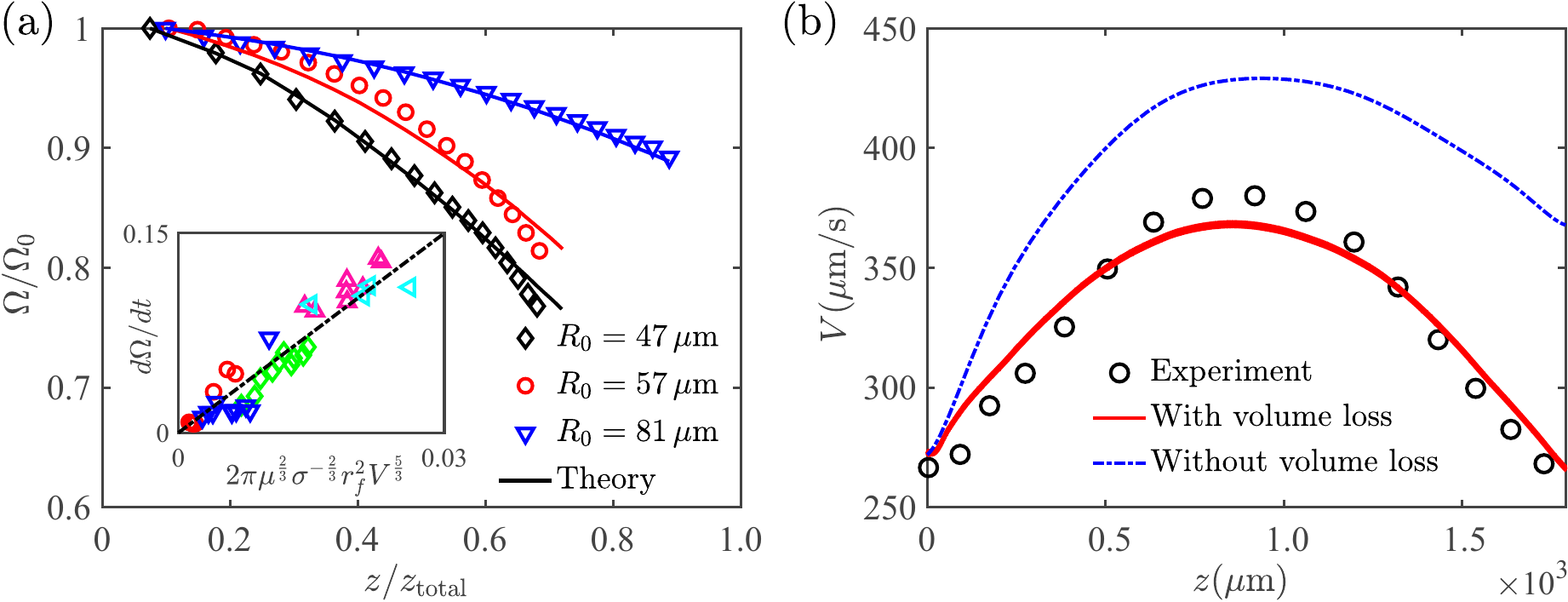}
\caption{\label{fig:loss} (a) The dimensionless volume $\Omega/\Omega_0$ as a function of migration distance $z/z_{\rm total}$ for droplets moving on a dry fiber. Theoretical lines are calculated according to Eq.\,(\ref{eq.loss}) with $C_w =5$. The inset shows the experimental temporal rate of volume loss $d\Omega/dt$ on a dry fiber, where the slope of the fitted dashed line is equal to the wetting factor $C_w$. (b) Velocity of a droplet migrating on a dry conical fiber, including experimental and theoretical results whether considering the droplet volume loss or not. $R_0=83~\rm{\mu m}$.}
\end{figure}

Subsequently, we explore whether the phenomenon observed on dry fibers significantly affects droplet volumes and migration. Since the film is too thin to measure directly, we track the droplet volume $\Omega$ as a function of the position $z$ instead, with the ratio of residual volumes $\Omega/\Omega_0$ shown in Fig.\,\ref{fig:loss}(a). A significant volume loss is observed, especially for smaller droplets, with a maximum loss ratio of up to 25\% over the entire migration. In contrast, the droplet volume on a prewetted fiber remains nearly constant. This effect could be explained by the Landau-Levich-Derjaguin theory, which examines the coating of a solid slowly pulled out of a liquid bath \cite{coating}. For fibers, the thickness $\delta$ of the coating film is given by 
\begin{equation}
    \delta=C_w r_f Ca^{2/3},
\end{equation}
where $Ca=\mu V/\sigma$ is the capillary number, and $C_w$ is a wetting factor depending on the liquid structure \cite{gravity-slide}. The temporal rate of the droplet volume loss can be expressed as
\begin{equation}
 \frac{d\Omega}{dt}=2\pi r_f\delta V=C_w2\pi r_f^2\mu^{2/3}\sigma^{-2/3}V^{5/3}.
    \label{eq.loss}
\end{equation}
The experimental volume loss is fitted according to Eq.\,(\ref{eq.loss}), yielding $C_w\approx 5$ as shown in the inset of Fig.\,\ref{fig:loss}(a).
Incorporating Eq.\,(\ref{eq.loss}) into the dynamic model, as illustrated in Fig.\,\ref{fig:loss}(b), provides a more accurate alignment with experimental data compared to predictions that ignore volume loss. This underscores the necessity of considering volume loss in modeling droplet migration on dry fibers.

\begin{figure}
\includegraphics[width=0.65\linewidth]{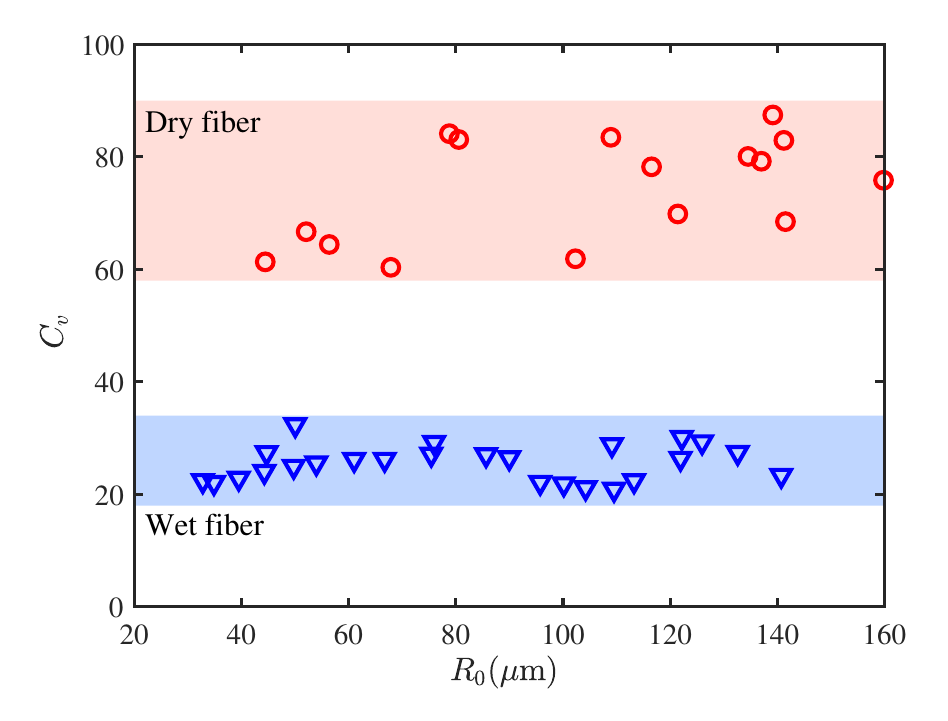}
\caption{\label{fig:cv} Friction coefficient $C_v$ fitted from different experimental cases as a function of droplet radius $R_0$. Red symbols for dry fibers and blue symbols for wet fibers.}
\end{figure}

Interestingly, an increase in the friction coefficient $C_v$ is observed when applying the dynamic model to describe droplet migration on dry fibers. This is reasonable as the precursor film effectively reduces the friction between the solid and liquid.
Figure\,\ref{fig:cv} presents the fitted $C_v$ as a function of droplet size $R_0$ from various experimental cases on dry (red symbols) and wet (blue symbols) fibers, with distinct regions for each condition. The data of droplets on dry fibers exhibit a wider range, primarily due to variations in microstructures among dry fibers, which can disrupt droplet migration, while a prewet film tends to reduce this variability. 
Notably, $C_v$ stabilizes after one or two droplets have migrated along an initially dry fiber, suggesting the fiber reaches a prewetted state covered by a stable liquid film.
We find $C_v^{\rm dry}\approx 75$ on dry fibers and $C_v^{\rm wet}\approx 25$ on prewetted fibers for silicone oil. As determined in Sec.\,\ref{sec.fv}, a minimal cutoff length $l_{\rm min}^{\rm dry}~\sim O(10^{-8}\,{\rm m})$ is deduced for dry fibers, which approximates the molecular size of silicone oil, while $l_{\rm min}^{\rm wet}~\sim O(10^{-5}\,{\rm m})$ for wet fibers, which is of the same order as the film thickness.

\section{\label{sec.con} conclusions}
In this article, we present an experimental investigation into the migration of barrel-shaped silicone oil droplets on glass conical fibers, with a particular focus on the effects of fiber geometry and surface conditions.
A dynamic model is proposed to describe the migration, incorporating a dimensionless coefficient $C_v$ to quantify the solid-liquid friction. 
Our findings indicate that the diverging geometry of conical fibers induces distinct acceleration and deceleration patterns in velocity profiles, while small cone angles enhance velocity differences when droplets move along or against gravity. Moreover, during migration on dry fibers, droplets form a thin film at the receding contact line, resulting in significant volume loss and reduced velocity, unlike the behavior observed on prewetted fibers. The dynamic model for dry migration is refined to include theoretical volume loss and an increased friction coefficient, better capturing these phenomena and physics.

In summary, this work provides a thorough and efficient framework for understanding how droplet motion is influenced by the properties of fibers and droplets. The findings can inform the design of fiber geometries to achieve specific objectives, such as directing droplets toward or away from a target and controlling droplet volume within a specific time, with potential applications in developing fiber arrays and networks.
Additionally, the model demonstrates the potential to accurately predict the hanging position or climbing height of large-scale droplets on fibers. This capability could offer significant theoretical insights for liquid manipulation and additive manufacturing, especially in space environments.

\section{ACKNOWLEDGMENTS}
This work was supported by National Key Research and Development Program of China (grant no. 2023YFB4603701), the National Natural Science Foundation of China (grant no. 12202437, 12388101, 12272372), Opening fund of State Key Laboratory of Nonlinear Mechanics, the Youth Innovation Promotion Association CAS (grant no. 2018491, 2023477), the Chinese Academy of Sciences Project for Young Scientists in Basic Research (grant no. YSBR-087) and the Fundamental Research Funds for the Central Universities (grant no. WK2090000051).

\appendix
\section{\label{app.size}Effects of droplet size}

\begin{figure}[h]
	\includegraphics[width=1\linewidth]{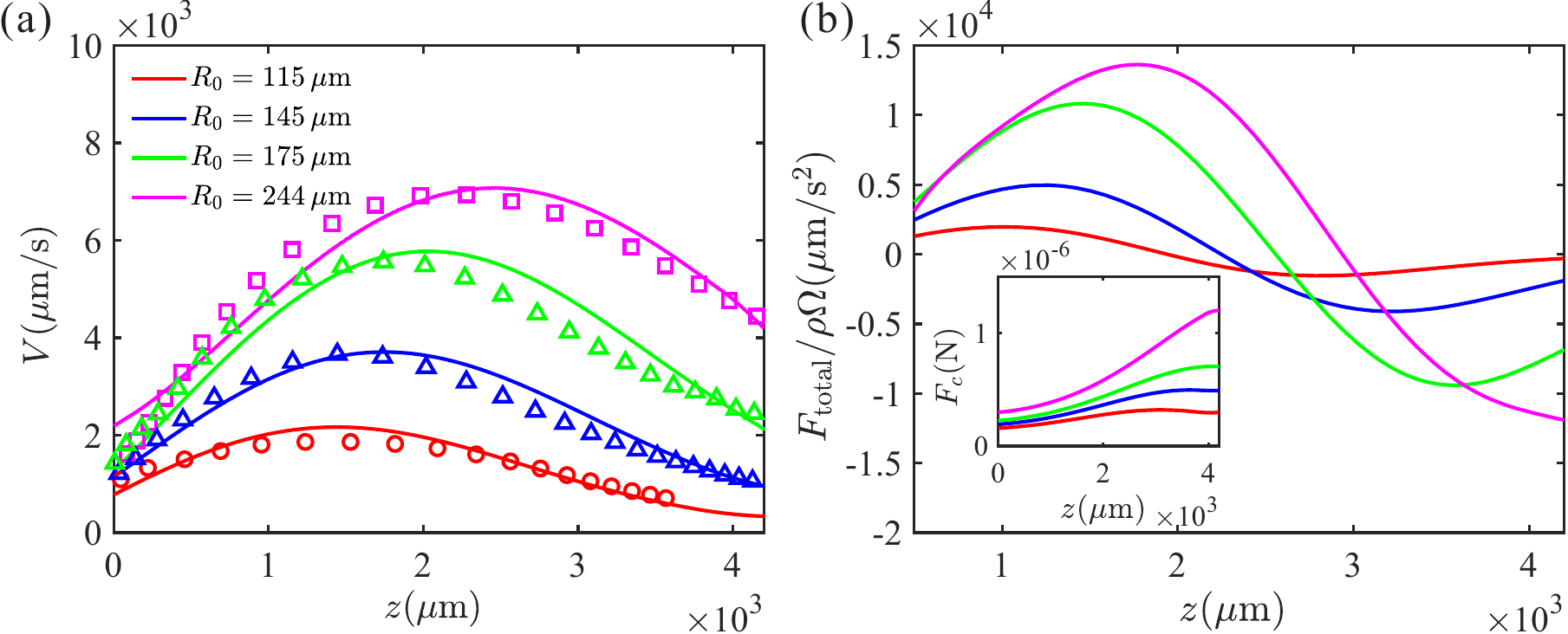}
	\caption{\label{fig:vfca} (a) Comparison of velocities between experimental results (symbols) and the dynamic model (curves); (b) Theoretical predictions of $F_c$ and total acceleration $F_{\rm total}/\rho\Omega$ of droplets with different size $R_0$.}
\end{figure}

Figure\,\ref{fig:vfca}\,(a) illustrates the impact of droplet sizes on velocity. Larger droplets experience more rapid acceleration and deceleration over longer distances, resulting in a higher maximum velocity. 
The inset of Fig.\,\ref{fig:vfca}\,(b) indicates that the driving force $F_c$ increases with droplet size $R_0$, attributed to the significant growth in droplet volume $\Omega$.
However, the variations in the total force $F_{\rm total}$, depicted in Fig.\,\ref{fig:vfca}\,(b), present a more complex behavior.

\section{\label{app.error}Estimation of errors due to experimental uncertainties in droplet size}

\begin{figure}[h]
	\includegraphics[width=0.5\linewidth]{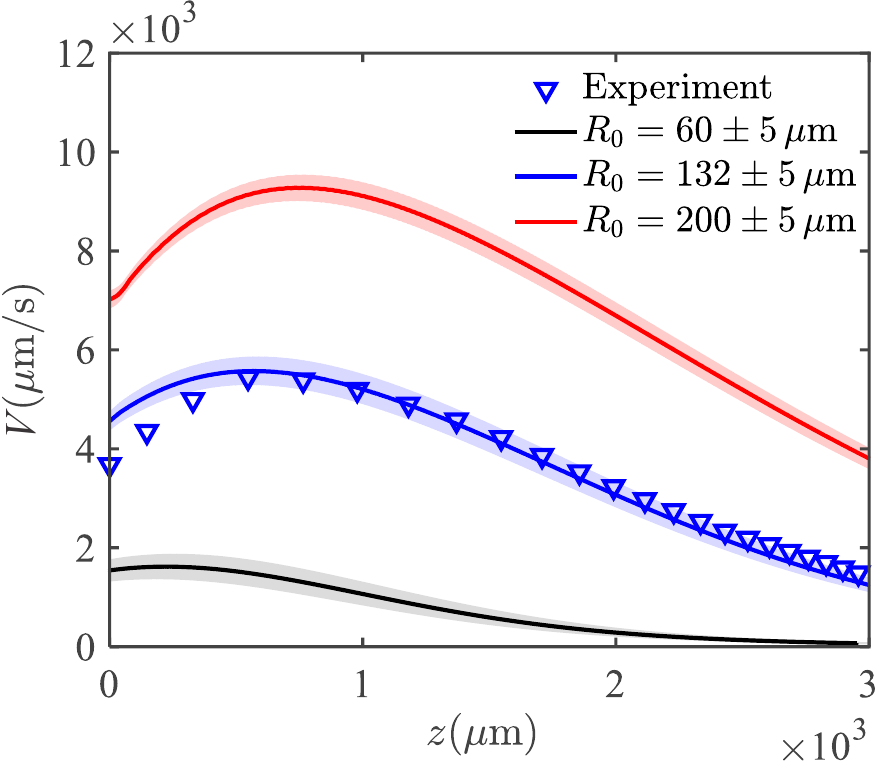}
	\caption{\label{fig:small_volume} The theoretically-predicted velocity of droplets with different sizes $R_0$ within a $\pm\,5\,\rm\mu m$ error on Cone 3.}
\end{figure}

In our experiments, droplet size is controlled by stopping infusion and allowing spontaneous pinch-off, influenced by the top opening geometry, liquid pump configuration and other perturbations. This method enabled size regulation within $\rm\pm2\,\mu m$. To access the impact of size fluctuations, we analyze the migration on Cone 3 in Fig.\,\ref{fig:cone}(b). Figure\,\ref{fig:small_volume} illustrates that small variations in droplet size, within $\rm\pm5\,\mu m$ (as indicated by the shaded area), can be negligible when considering the influence of cone shape on droplet dynamics, especially compared to the experimental curve.

\section{\label{app.g}Effects of gravity on the droplet shape}

\begin{figure}[h]
	\includegraphics[width=1\linewidth]{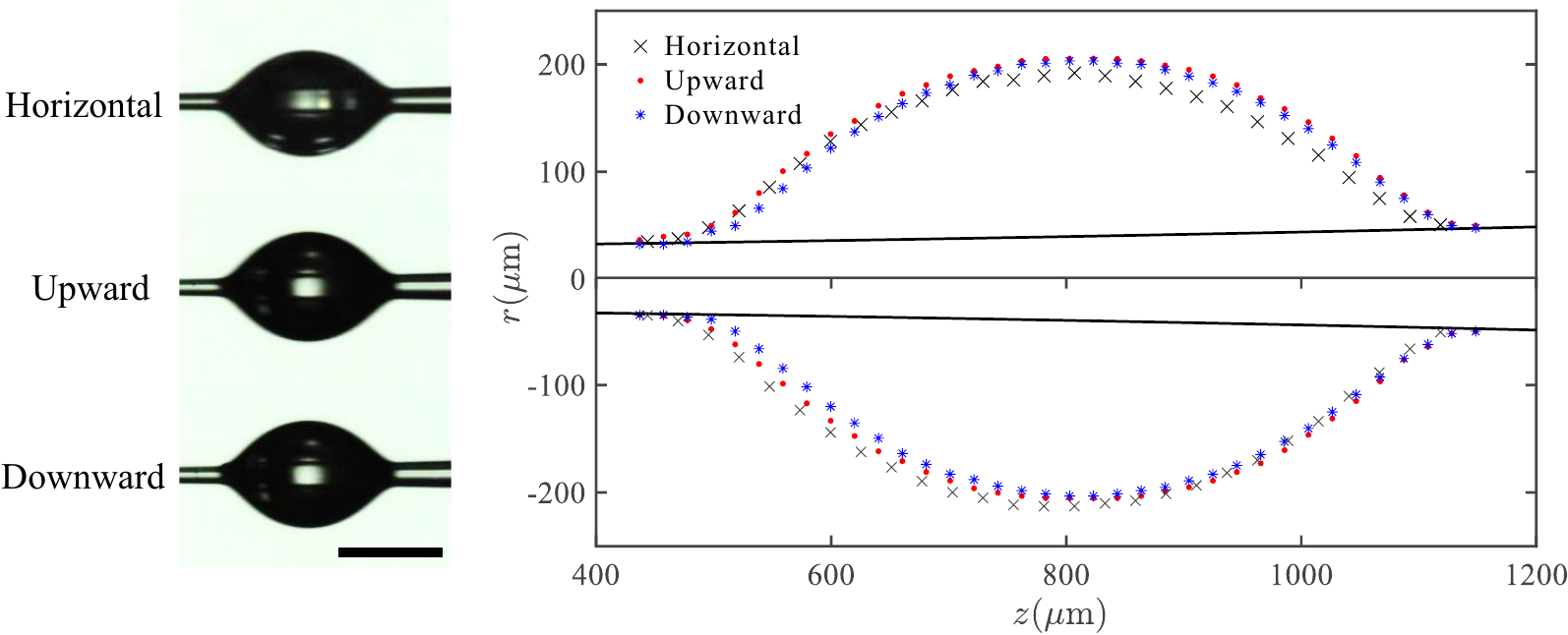}
	\caption{\label{fig:g-shape} Shape comparison of droplets at the same position on a fiber placed horizontally, vertically upwards and downwards. $R_0\approx220\,\rm{\mu m}$. Scale bar: $500\,\rm{\mu m}$.}
\end{figure}

As the droplet size increases, its center of mass on a horizontal fiber shifts downward, causing it to deviate from the fiber axis under the influence of gravitational forces. This deviation leads to a loss of rotational symmetry around the fiber axis in the droplet's shape. The degree of deformation is influenced by the Bond number $Bo=\rho gr_f^2/\sigma$ and the dimensionless volume $\Omega^*=\Omega/r_f^3$, with larger values of $Bo$ and $\Omega^*$ resulting in more significant deviations \cite{gravity-shape}.
We compare the symmetry of droplets on fibers with different orientations in Fig.\,\ref{fig:g-shape}. 
In our experiments, droplets with $R_0\approx220\,\rm{\mu m}$ exhibit deviations of less than $10\%$ between the upper and lower surfaces, while droplet shapes on upward- and downward-pointing fibers show negligible differences, indicating minimal gravitational effects on droplet curvatures and driving forces in our model.

\bibliography{bibtex}

\providecommand{\noopsort}[1]{}\providecommand{\singleletter}[1]{#1}%
\begin{thebibliography}{34}%
\makeatletter
\providecommand \@ifxundefined [1]{%
 \@ifx{#1\undefined}
}%
\providecommand \@ifnum [1]{%
 \ifnum #1\expandafter \@firstoftwo
 \else \expandafter \@secondoftwo
 \fi
}%
\providecommand \@ifx [1]{%
 \ifx #1\expandafter \@firstoftwo
 \else \expandafter \@secondoftwo
 \fi
}%
\providecommand \natexlab [1]{#1}%
\providecommand \enquote  [1]{``#1''}%
\providecommand \bibnamefont  [1]{#1}%
\providecommand \bibfnamefont [1]{#1}%
\providecommand \citenamefont [1]{#1}%
\providecommand \href@noop [0]{\@secondoftwo}%
\providecommand \href [0]{\begingroup \@sanitize@url \@href}%
\providecommand \@href[1]{\@@startlink{#1}\@@href}%
\providecommand \@@href[1]{\endgroup#1\@@endlink}%
\providecommand \@sanitize@url [0]{\catcode `\\12\catcode `\$12\catcode `\&12\catcode `\#12\catcode `\^12\catcode `\_12\catcode `\%12\relax}%
\providecommand \@@startlink[1]{}%
\providecommand \@@endlink[0]{}%
\providecommand \url  [0]{\begingroup\@sanitize@url \@url }%
\providecommand \@url [1]{\endgroup\@href {#1}{\urlprefix }}%
\providecommand \urlprefix  [0]{URL }%
\providecommand \Eprint [0]{\href }%
\providecommand \doibase [0]{https://doi.org/}%
\providecommand \selectlanguage [0]{\@gobble}%
\providecommand \bibinfo  [0]{\@secondoftwo}%
\providecommand \bibfield  [0]{\@secondoftwo}%
\providecommand \translation [1]{[#1]}%
\providecommand \BibitemOpen [0]{}%
\providecommand \bibitemStop [0]{}%
\providecommand \bibitemNoStop [0]{.\EOS\space}%
\providecommand \EOS [0]{\spacefactor3000\relax}%
\providecommand \BibitemShut  [1]{\csname bibitem#1\endcsname}%
\let\auto@bib@innerbib\@empty
\bibitem [{\citenamefont {Zheng}\ \emph {et~al.}(2010)\citenamefont {Zheng}, \citenamefont {Bai}, \citenamefont {Huang}, \citenamefont {Tian}, \citenamefont {Nie}, \citenamefont {Zhao}, \citenamefont {Zhai},\ and\ \citenamefont {Jiang}}]{spider}%
  \BibitemOpen
  \bibfield  {author} {\bibinfo {author} {\bibfnamefont {Y.}~\bibnamefont {Zheng}}, \bibinfo {author} {\bibfnamefont {H.}~\bibnamefont {Bai}}, \bibinfo {author} {\bibfnamefont {Z.}~\bibnamefont {Huang}}, \bibinfo {author} {\bibfnamefont {X.}~\bibnamefont {Tian}}, \bibinfo {author} {\bibfnamefont {F.~Q.}\ \bibnamefont {Nie}}, \bibinfo {author} {\bibfnamefont {Y.}~\bibnamefont {Zhao}}, \bibinfo {author} {\bibfnamefont {J.}~\bibnamefont {Zhai}},\ and\ \bibinfo {author} {\bibfnamefont {L.}~\bibnamefont {Jiang}},\ }\bibfield  {title} {\enquote {\bibinfo {title} {Directional water collection on wetted spider silk},}\ }\href@noop {} {\bibfield  {journal} {\bibinfo  {journal} {Nature}\ }\textbf {\bibinfo {volume} {463}},\ \bibinfo {pages} {640--3} (\bibinfo {year} {2010})}\BibitemShut {NoStop}%
\bibitem [{\citenamefont {Ju}\ \emph {et~al.}(2012)\citenamefont {Ju}, \citenamefont {Bai}, \citenamefont {Zheng}, \citenamefont {Zhao}, \citenamefont {Fang},\ and\ \citenamefont {Jiang}}]{catus}%
  \BibitemOpen
  \bibfield  {author} {\bibinfo {author} {\bibfnamefont {J.}~\bibnamefont {Ju}}, \bibinfo {author} {\bibfnamefont {H.}~\bibnamefont {Bai}}, \bibinfo {author} {\bibfnamefont {Y.}~\bibnamefont {Zheng}}, \bibinfo {author} {\bibfnamefont {T.}~\bibnamefont {Zhao}}, \bibinfo {author} {\bibfnamefont {R.}~\bibnamefont {Fang}},\ and\ \bibinfo {author} {\bibfnamefont {L.}~\bibnamefont {Jiang}},\ }\bibfield  {title} {\enquote {\bibinfo {title} {A multi-structural and multi-functional integrated fog collection system in cactus},}\ }\href@noop {} {\bibfield  {journal} {\bibinfo  {journal} {Nat. Commun.}\ }\textbf {\bibinfo {volume} {3}},\ \bibinfo {pages} {1247} (\bibinfo {year} {2012})}\BibitemShut {NoStop}%
\bibitem [{\citenamefont {Wang}\ \emph {et~al.}(2015)\citenamefont {Wang}, \citenamefont {Yao}, \citenamefont {Liu}, \citenamefont {Quere},\ and\ \citenamefont {Jiang}}]{Self-removal}%
  \BibitemOpen
  \bibfield  {author} {\bibinfo {author} {\bibfnamefont {Q.}~\bibnamefont {Wang}}, \bibinfo {author} {\bibfnamefont {X.}~\bibnamefont {Yao}}, \bibinfo {author} {\bibfnamefont {H.}~\bibnamefont {Liu}}, \bibinfo {author} {\bibfnamefont {D.}~\bibnamefont {Quere}},\ and\ \bibinfo {author} {\bibfnamefont {L.}~\bibnamefont {Jiang}},\ }\bibfield  {title} {\enquote {\bibinfo {title} {Self-removal of condensed water on the legs of water striders},}\ }\href@noop {} {\bibfield  {journal} {\bibinfo  {journal} {Proc. Natl. Acad. Sci. U S A}\ }\textbf {\bibinfo {volume} {112}},\ \bibinfo {pages} {9247--52} (\bibinfo {year} {2015})}\BibitemShut {NoStop}%
\bibitem [{\citenamefont {Zhang}\ \emph {et~al.}(2023)\citenamefont {Zhang}, \citenamefont {Wang}, \citenamefont {Liu}, \citenamefont {Gao}, \citenamefont {Song}, \citenamefont {Cheng}, \citenamefont {Zhang},\ and\ \citenamefont {Ding}}]{cross-cone}%
  \BibitemOpen
  \bibfield  {author} {\bibinfo {author} {\bibfnamefont {F.}~\bibnamefont {Zhang}}, \bibinfo {author} {\bibfnamefont {Z.}~\bibnamefont {Wang}}, \bibinfo {author} {\bibfnamefont {Z.}~\bibnamefont {Liu}}, \bibinfo {author} {\bibfnamefont {X.}~\bibnamefont {Gao}}, \bibinfo {author} {\bibfnamefont {Y.}~\bibnamefont {Song}}, \bibinfo {author} {\bibfnamefont {G.}~\bibnamefont {Cheng}}, \bibinfo {author} {\bibfnamefont {Z.}~\bibnamefont {Zhang}},\ and\ \bibinfo {author} {\bibfnamefont {J.}~\bibnamefont {Ding}},\ }\bibfield  {title} {\enquote {\bibinfo {title} {Cross-hatch textured cone enables dual-mode water transport and collection},}\ }\href@noop {} {\bibfield  {journal} {\bibinfo  {journal} {Chem. Eng. J.}\ }\textbf {\bibinfo {volume} {478}},\ \bibinfo {pages} {147336} (\bibinfo {year} {2023})}\BibitemShut {NoStop}%
\bibitem [{\citenamefont {Hafley}, \citenamefont {Taminger},\ and\ \citenamefont {Bird}(2007)}]{Electron}%
  \BibitemOpen
  \bibfield  {author} {\bibinfo {author} {\bibfnamefont {R.}~\bibnamefont {Hafley}}, \bibinfo {author} {\bibfnamefont {K.}~\bibnamefont {Taminger}},\ and\ \bibinfo {author} {\bibfnamefont {R.}~\bibnamefont {Bird}},\ }\bibfield  {title} {\enquote {\bibinfo {title} {Electron beam freeform fabrication in the space environment},}\ }in\ \href@noop {} {\emph {\bibinfo {booktitle} {45th AIAA Aerospace Sciences Meeting and Exhibit}}}\ (\bibinfo {address} {Reno, NV, USA},\ \bibinfo {year} {2007})\BibitemShut {NoStop}%
\bibitem [{\citenamefont {Wu}\ \emph {et~al.}(2022)\citenamefont {Wu}, \citenamefont {Zhu}, \citenamefont {Wu}, \citenamefont {Xie}, \citenamefont {Qian}, \citenamefont {Yin},\ and\ \citenamefont {Yang}}]{liquidmetal}%
  \BibitemOpen
  \bibfield  {author} {\bibinfo {author} {\bibfnamefont {Q.}~\bibnamefont {Wu}}, \bibinfo {author} {\bibfnamefont {F.}~\bibnamefont {Zhu}}, \bibinfo {author} {\bibfnamefont {Z.}~\bibnamefont {Wu}}, \bibinfo {author} {\bibfnamefont {Y.}~\bibnamefont {Xie}}, \bibinfo {author} {\bibfnamefont {J.}~\bibnamefont {Qian}}, \bibinfo {author} {\bibfnamefont {J.}~\bibnamefont {Yin}},\ and\ \bibinfo {author} {\bibfnamefont {H.}~\bibnamefont {Yang}},\ }\bibfield  {title} {\enquote {\bibinfo {title} {Suspension printing of liquid metal in yield-stress fluid for resilient 3d constructs with electromagnetic functions},}\ }\href@noop {} {\bibfield  {journal} {\bibinfo  {journal} {npj Flex. Electron.}\ }\textbf {\bibinfo {volume} {6}},\ \bibinfo {pages} {50} (\bibinfo {year} {2022})}\BibitemShut {NoStop}%
\bibitem [{\citenamefont {Gilet}, \citenamefont {Terwagne},\ and\ \citenamefont {Vandewalle}(2010)}]{gravity-slide}%
  \BibitemOpen
  \bibfield  {author} {\bibinfo {author} {\bibfnamefont {T.}~\bibnamefont {Gilet}}, \bibinfo {author} {\bibfnamefont {D.}~\bibnamefont {Terwagne}},\ and\ \bibinfo {author} {\bibfnamefont {N.}~\bibnamefont {Vandewalle}},\ }\bibfield  {title} {\enquote {\bibinfo {title} {Droplets sliding on fibres},}\ }\href@noop {} {\bibfield  {journal} {\bibinfo  {journal} {Eur. Phys. J. E. Soft Matter}\ }\textbf {\bibinfo {volume} {31}},\ \bibinfo {pages} {253--62} (\bibinfo {year} {2010})}\BibitemShut {NoStop}%
\bibitem [{\citenamefont {Christianto}\ \emph {et~al.}(2022)\citenamefont {Christianto}, \citenamefont {Rahmawan}, \citenamefont {Semprebon},\ and\ \citenamefont {Kusumaatmaja}}]{gravity2}%
  \BibitemOpen
  \bibfield  {author} {\bibinfo {author} {\bibfnamefont {R.}~\bibnamefont {Christianto}}, \bibinfo {author} {\bibfnamefont {Y.}~\bibnamefont {Rahmawan}}, \bibinfo {author} {\bibfnamefont {C.}~\bibnamefont {Semprebon}},\ and\ \bibinfo {author} {\bibfnamefont {H.}~\bibnamefont {Kusumaatmaja}},\ }\bibfield  {title} {\enquote {\bibinfo {title} {Modeling the dynamics of partially wetting droplets on fibers},}\ }\href@noop {} {\bibfield  {journal} {\bibinfo  {journal} {Phys. Rev. Fluids}\ }\textbf {\bibinfo {volume} {7}},\ \bibinfo {pages} {103606} (\bibinfo {year} {2022})}\BibitemShut {NoStop}%
\bibitem [{\citenamefont {Poulain}\ and\ \citenamefont {Carlson}(2023)}]{vibrate}%
  \BibitemOpen
  \bibfield  {author} {\bibinfo {author} {\bibfnamefont {S.}~\bibnamefont {Poulain}}\ and\ \bibinfo {author} {\bibfnamefont {A.}~\bibnamefont {Carlson}},\ }\bibfield  {title} {\enquote {\bibinfo {title} {Sliding, vibrating and swinging droplets on an oscillating fibre},}\ }\href@noop {} {\bibfield  {journal} {\bibinfo  {journal} {J. Fluid Mech.}\ }\textbf {\bibinfo {volume} {967}},\ \bibinfo {pages} {A24} (\bibinfo {year} {2023})}\BibitemShut {NoStop}%
\bibitem [{\citenamefont {Corpart}, \citenamefont {Restagno},\ and\ \citenamefont {Boulogne}(2023)}]{coffee}%
  \BibitemOpen
  \bibfield  {author} {\bibinfo {author} {\bibfnamefont {M.}~\bibnamefont {Corpart}}, \bibinfo {author} {\bibfnamefont {F.}~\bibnamefont {Restagno}},\ and\ \bibinfo {author} {\bibfnamefont {F.}~\bibnamefont {Boulogne}},\ }\bibfield  {title} {\enquote {\bibinfo {title} {Coffee stain effect on a fibre from axisymmetric droplets},}\ }\href@noop {} {\bibfield  {journal} {\bibinfo  {journal} {J. Fluid Mech.}\ }\textbf {\bibinfo {volume} {957}},\ \bibinfo {pages} {A24} (\bibinfo {year} {2023})}\BibitemShut {NoStop}%
\bibitem [{\citenamefont {Bintein}\ \emph {et~al.}(2019)\citenamefont {Bintein}, \citenamefont {Bense}, \citenamefont {Clanet},\ and\ \citenamefont {Quéré}}]{crosswind}%
  \BibitemOpen
  \bibfield  {author} {\bibinfo {author} {\bibfnamefont {P.-B.}\ \bibnamefont {Bintein}}, \bibinfo {author} {\bibfnamefont {H.}~\bibnamefont {Bense}}, \bibinfo {author} {\bibfnamefont {C.}~\bibnamefont {Clanet}},\ and\ \bibinfo {author} {\bibfnamefont {D.}~\bibnamefont {Quéré}},\ }\bibfield  {title} {\enquote {\bibinfo {title} {Self-propelling droplets on fibres subject to a crosswind},}\ }\href@noop {} {\bibfield  {journal} {\bibinfo  {journal} {Nat. Phys.}\ }\textbf {\bibinfo {volume} {15}},\ \bibinfo {pages} {1027--1032} (\bibinfo {year} {2019})}\BibitemShut {NoStop}%
\bibitem [{\citenamefont {Weislogel}(1997)}]{theta-induce}%
  \BibitemOpen
  \bibfield  {author} {\bibinfo {author} {\bibfnamefont {M.~M.}\ \bibnamefont {Weislogel}},\ }\bibfield  {title} {\enquote {\bibinfo {title} {Steady spontaneous capillary flow in partially coated tubes},}\ }\href@noop {} {\bibfield  {journal} {\bibinfo  {journal} {AIChE J.}\ }\textbf {\bibinfo {volume} {43}},\ \bibinfo {pages} {645--654} (\bibinfo {year} {1997})}\BibitemShut {NoStop}%
\bibitem [{\citenamefont {Ju}\ \emph {et~al.}(2013)\citenamefont {Ju}, \citenamefont {Xiao}, \citenamefont {Yao}, \citenamefont {Bai},\ and\ \citenamefont {Jiang}}]{fog-catus}%
  \BibitemOpen
  \bibfield  {author} {\bibinfo {author} {\bibfnamefont {J.}~\bibnamefont {Ju}}, \bibinfo {author} {\bibfnamefont {K.}~\bibnamefont {Xiao}}, \bibinfo {author} {\bibfnamefont {X.}~\bibnamefont {Yao}}, \bibinfo {author} {\bibfnamefont {H.}~\bibnamefont {Bai}},\ and\ \bibinfo {author} {\bibfnamefont {L.}~\bibnamefont {Jiang}},\ }\bibfield  {title} {\enquote {\bibinfo {title} {Bioinspired conical copper wire with gradient wettability for continuous and efficient fog collection},}\ }\href@noop {} {\bibfield  {journal} {\bibinfo  {journal} {Adv. Mater.}\ }\textbf {\bibinfo {volume} {25}},\ \bibinfo {pages} {5937--42} (\bibinfo {year} {2013})}\BibitemShut {NoStop}%
\bibitem [{\citenamefont {Liu}\ \emph {et~al.}(2015)\citenamefont {Liu}, \citenamefont {Xue}, \citenamefont {Chen},\ and\ \citenamefont {Zheng}}]{mirco-structure}%
  \BibitemOpen
  \bibfield  {author} {\bibinfo {author} {\bibfnamefont {C.}~\bibnamefont {Liu}}, \bibinfo {author} {\bibfnamefont {Y.}~\bibnamefont {Xue}}, \bibinfo {author} {\bibfnamefont {Y.}~\bibnamefont {Chen}},\ and\ \bibinfo {author} {\bibfnamefont {Y.}~\bibnamefont {Zheng}},\ }\bibfield  {title} {\enquote {\bibinfo {title} {Effective directional self-gathering of drops on spine of cactus with splayed capillary arrays},}\ }\href@noop {} {\bibfield  {journal} {\bibinfo  {journal} {Sci. Rep.}\ }\textbf {\bibinfo {volume} {5}},\ \bibinfo {pages} {17757} (\bibinfo {year} {2015})}\BibitemShut {NoStop}%
\bibitem [{\citenamefont {Tan}\ \emph {et~al.}(2016)\citenamefont {Tan}, \citenamefont {Zhu}, \citenamefont {Shi}, \citenamefont {Tang},\ and\ \citenamefont {Liao}}]{micro-stru2}%
  \BibitemOpen
  \bibfield  {author} {\bibinfo {author} {\bibfnamefont {X.}~\bibnamefont {Tan}}, \bibinfo {author} {\bibfnamefont {Y.}~\bibnamefont {Zhu}}, \bibinfo {author} {\bibfnamefont {T.}~\bibnamefont {Shi}}, \bibinfo {author} {\bibfnamefont {Z.}~\bibnamefont {Tang}},\ and\ \bibinfo {author} {\bibfnamefont {G.}~\bibnamefont {Liao}},\ }\bibfield  {title} {\enquote {\bibinfo {title} {Patterned gradient surface for spontaneous droplet transportation and water collection: simulation and experiment},}\ }\href@noop {} {\bibfield  {journal} {\bibinfo  {journal} {J. Micromech. Microeng.}\ }\textbf {\bibinfo {volume} {26}},\ \bibinfo {pages} {115009} (\bibinfo {year} {2016})}\BibitemShut {NoStop}%
\bibitem [{\citenamefont {Yang}\ \emph {et~al.}(2008)\citenamefont {Yang}, \citenamefont {Yang}, \citenamefont {Chen},\ and\ \citenamefont {Yao}}]{micropatterned}%
  \BibitemOpen
  \bibfield  {author} {\bibinfo {author} {\bibfnamefont {J.~T.}\ \bibnamefont {Yang}}, \bibinfo {author} {\bibfnamefont {Z.~H.}\ \bibnamefont {Yang}}, \bibinfo {author} {\bibfnamefont {C.~Y.}\ \bibnamefont {Chen}},\ and\ \bibinfo {author} {\bibfnamefont {D.~J.}\ \bibnamefont {Yao}},\ }\bibfield  {title} {\enquote {\bibinfo {title} {Conversion of surface energy and manipulation of a single droplet across micropatterned surfaces},}\ }\href@noop {} {\bibfield  {journal} {\bibinfo  {journal} {Langmuir}\ }\textbf {\bibinfo {volume} {24}},\ \bibinfo {pages} {9889--97} (\bibinfo {year} {2008})}\BibitemShut {NoStop}%
\bibitem [{\citenamefont {Carroll}(1986)}]{Carroll-rollup}%
  \BibitemOpen
  \bibfield  {author} {\bibinfo {author} {\bibfnamefont {B.~J.}\ \bibnamefont {Carroll}},\ }\bibfield  {title} {\enquote {\bibinfo {title} {Equilibrium conformations of liquid drops on thin cylinders under forces of capillarity. a theory for the roll-up process},}\ }\href@noop {} {\bibfield  {journal} {\bibinfo  {journal} {Langmuir}\ }\textbf {\bibinfo {volume} {2}},\ \bibinfo {pages} {248--250} (\bibinfo {year} {1986})}\BibitemShut {NoStop}%
\bibitem [{\citenamefont {Carroll}(1976)}]{carroll-barrel}%
  \BibitemOpen
  \bibfield  {author} {\bibinfo {author} {\bibfnamefont {B.~J.}\ \bibnamefont {Carroll}},\ }\bibfield  {title} {\enquote {\bibinfo {title} {The accurate measurement of contact angle, phase contact areas, drop volume, and laplace excess pressure in drop-on-fiber systems},}\ }\href@noop {} {\bibfield  {journal} {\bibinfo  {journal} {J. Colloid Interface Sci.}\ }\textbf {\bibinfo {volume} {57}},\ \bibinfo {pages} {488--495} (\bibinfo {year} {1976})}\BibitemShut {NoStop}%
\bibitem [{\citenamefont {Lorenceau}\ and\ \citenamefont {Quere}(2004)}]{lq-wire}%
  \BibitemOpen
  \bibfield  {author} {\bibinfo {author} {\bibfnamefont {E.}~\bibnamefont {Lorenceau}}\ and\ \bibinfo {author} {\bibfnamefont {D.}~\bibnamefont {Quere}},\ }\bibfield  {title} {\enquote {\bibinfo {title} {Drops on a conical wire},}\ }\href@noop {} {\bibfield  {journal} {\bibinfo  {journal} {J. Fluid Mech.}\ }\textbf {\bibinfo {volume} {510}},\ \bibinfo {pages} {29--45} (\bibinfo {year} {2004})}\BibitemShut {NoStop}%
\bibitem [{\citenamefont {Li}\ and\ \citenamefont {Thoroddsen}(2013)}]{li-fiber}%
  \BibitemOpen
  \bibfield  {author} {\bibinfo {author} {\bibfnamefont {E.~Q.}\ \bibnamefont {Li}}\ and\ \bibinfo {author} {\bibfnamefont {S.~T.}\ \bibnamefont {Thoroddsen}},\ }\bibfield  {title} {\enquote {\bibinfo {title} {The fastest drop climbing on a wet conical fibre},}\ }\href@noop {} {\bibfield  {journal} {\bibinfo  {journal} {Phys. Fluids}\ }\textbf {\bibinfo {volume} {25}},\ \bibinfo {pages} {052105} (\bibinfo {year} {2013})}\BibitemShut {NoStop}%
\bibitem [{\citenamefont {Fournier}\ \emph {et~al.}(2021)\citenamefont {Fournier}, \citenamefont {Lee}, \citenamefont {Schulman}, \citenamefont {Raphael},\ and\ \citenamefont {Dalnoki-Veress}}]{law_f}%
  \BibitemOpen
  \bibfield  {author} {\bibinfo {author} {\bibfnamefont {C.}~\bibnamefont {Fournier}}, \bibinfo {author} {\bibfnamefont {C.~L.}\ \bibnamefont {Lee}}, \bibinfo {author} {\bibfnamefont {R.~D.}\ \bibnamefont {Schulman}}, \bibinfo {author} {\bibfnamefont {E.}~\bibnamefont {Raphael}},\ and\ \bibinfo {author} {\bibfnamefont {K.}~\bibnamefont {Dalnoki-Veress}},\ }\bibfield  {title} {\enquote {\bibinfo {title} {Droplet migration on conical fibers},}\ }\href@noop {} {\bibfield  {journal} {\bibinfo  {journal} {Eur. Phys. J. E}\ }\textbf {\bibinfo {volume} {44}},\ \bibinfo {pages} {12} (\bibinfo {year} {2021})}\BibitemShut {NoStop}%
\bibitem [{\citenamefont {Van~Hulle}\ \emph {et~al.}(2021)\citenamefont {Van~Hulle}, \citenamefont {Weyer}, \citenamefont {Dorbolo},\ and\ \citenamefont {Vandewalle}}]{law_v}%
  \BibitemOpen
  \bibfield  {author} {\bibinfo {author} {\bibfnamefont {J.}~\bibnamefont {Van~Hulle}}, \bibinfo {author} {\bibfnamefont {F.}~\bibnamefont {Weyer}}, \bibinfo {author} {\bibfnamefont {S.}~\bibnamefont {Dorbolo}},\ and\ \bibinfo {author} {\bibfnamefont {N.}~\bibnamefont {Vandewalle}},\ }\bibfield  {title} {\enquote {\bibinfo {title} {Capillary transport from barrel to clamshell droplets on conical fibers},}\ }\href@noop {} {\bibfield  {journal} {\bibinfo  {journal} {Phys. Rev. Fluids}\ }\textbf {\bibinfo {volume} {6}} (\bibinfo {year} {2021})}\BibitemShut {NoStop}%
\bibitem [{\citenamefont {Liu}\ \emph {et~al.}(2007)\citenamefont {Liu}, \citenamefont {Xia}, \citenamefont {Li},\ and\ \citenamefont {Feng}}]{surface2-tube}%
  \BibitemOpen
  \bibfield  {author} {\bibinfo {author} {\bibfnamefont {J.~L.}\ \bibnamefont {Liu}}, \bibinfo {author} {\bibfnamefont {R.}~\bibnamefont {Xia}}, \bibinfo {author} {\bibfnamefont {B.~W.}\ \bibnamefont {Li}},\ and\ \bibinfo {author} {\bibfnamefont {X.~Q.}\ \bibnamefont {Feng}},\ }\bibfield  {title} {\enquote {\bibinfo {title} {Directional motion of droplets in a conical tube or on a conical fibre},}\ }\href@noop {} {\bibfield  {journal} {\bibinfo  {journal} {Chin. Phys. Lett.}\ }\textbf {\bibinfo {volume} {24}},\ \bibinfo {pages} {3210--3213} (\bibinfo {year} {2007})}\BibitemShut {NoStop}%
\bibitem [{\citenamefont {Michielsen}\ \emph {et~al.}(2011)\citenamefont {Michielsen}, \citenamefont {Zhang}, \citenamefont {Du},\ and\ \citenamefont {Lee}}]{surface-ene3}%
  \BibitemOpen
  \bibfield  {author} {\bibinfo {author} {\bibfnamefont {S.}~\bibnamefont {Michielsen}}, \bibinfo {author} {\bibfnamefont {J.}~\bibnamefont {Zhang}}, \bibinfo {author} {\bibfnamefont {J.}~\bibnamefont {Du}},\ and\ \bibinfo {author} {\bibfnamefont {H.~J.}\ \bibnamefont {Lee}},\ }\bibfield  {title} {\enquote {\bibinfo {title} {Gibbs free energy of liquid drops on conical fibers},}\ }\href@noop {} {\bibfield  {journal} {\bibinfo  {journal} {Langmuir}\ }\textbf {\bibinfo {volume} {27}},\ \bibinfo {pages} {11867--72} (\bibinfo {year} {2011})}\BibitemShut {NoStop}%
\bibitem [{\citenamefont {Chan}, \citenamefont {Yang},\ and\ \citenamefont {Carlson}(2020)}]{chan-drop}%
  \BibitemOpen
  \bibfield  {author} {\bibinfo {author} {\bibfnamefont {T.~S.}\ \bibnamefont {Chan}}, \bibinfo {author} {\bibfnamefont {F.}~\bibnamefont {Yang}},\ and\ \bibinfo {author} {\bibfnamefont {A.}~\bibnamefont {Carlson}},\ }\bibfield  {title} {\enquote {\bibinfo {title} {Directional spreading of a viscous droplet on a conical fibre},}\ }\href@noop {} {\bibfield  {journal} {\bibinfo  {journal} {J. Fluid Mech.}\ }\textbf {\bibinfo {volume} {894}},\ \bibinfo {pages} {A26} (\bibinfo {year} {2020})}\BibitemShut {NoStop}%
\bibitem [{\citenamefont {Chan}\ \emph {et~al.}(2021{\natexlab{a}})\citenamefont {Chan}, \citenamefont {Lee}, \citenamefont {Pedersen}, \citenamefont {Dalnoki-Veress},\ and\ \citenamefont {Carlson}}]{chan-coating-prf}%
  \BibitemOpen
  \bibfield  {author} {\bibinfo {author} {\bibfnamefont {T.~S.}\ \bibnamefont {Chan}}, \bibinfo {author} {\bibfnamefont {C.~L.}\ \bibnamefont {Lee}}, \bibinfo {author} {\bibfnamefont {C.}~\bibnamefont {Pedersen}}, \bibinfo {author} {\bibfnamefont {K.}~\bibnamefont {Dalnoki-Veress}},\ and\ \bibinfo {author} {\bibfnamefont {A.}~\bibnamefont {Carlson}},\ }\bibfield  {title} {\enquote {\bibinfo {title} {Film coating by directional droplet spreading on fibers},}\ }\href@noop {} {\bibfield  {journal} {\bibinfo  {journal} {Phys. Rev. Fluids}\ }\textbf {\bibinfo {volume} {6}},\ \bibinfo {pages} {014004} (\bibinfo {year} {2021}{\natexlab{a}})}\BibitemShut {NoStop}%
\bibitem [{\citenamefont {Chan}\ \emph {et~al.}(2021{\natexlab{b}})\citenamefont {Chan}, \citenamefont {Pedersen}, \citenamefont {Koplik},\ and\ \citenamefont {Carlson}}]{chan-film}%
  \BibitemOpen
  \bibfield  {author} {\bibinfo {author} {\bibfnamefont {T.~S.}\ \bibnamefont {Chan}}, \bibinfo {author} {\bibfnamefont {C.}~\bibnamefont {Pedersen}}, \bibinfo {author} {\bibfnamefont {J.}~\bibnamefont {Koplik}},\ and\ \bibinfo {author} {\bibfnamefont {A.}~\bibnamefont {Carlson}},\ }\bibfield  {title} {\enquote {\bibinfo {title} {Film deposition and dynamics of a self-propelled wetting droplet on a conical fibre},}\ }\href@noop {} {\bibfield  {journal} {\bibinfo  {journal} {J. Fluid Mech.}\ }\textbf {\bibinfo {volume} {907}},\ \bibinfo {pages} {A29} (\bibinfo {year} {2021}{\natexlab{b}})}\BibitemShut {NoStop}%
\bibitem [{\citenamefont {de~Gennes}, \citenamefont {Brochard-Wyart},\ and\ \citenamefont {Quéré}(2004)}]{capillary-wetting}%
  \BibitemOpen
  \bibfield  {author} {\bibinfo {author} {\bibfnamefont {P.-G.}\ \bibnamefont {de~Gennes}}, \bibinfo {author} {\bibfnamefont {F.}~\bibnamefont {Brochard-Wyart}},\ and\ \bibinfo {author} {\bibfnamefont {D.}~\bibnamefont {Quéré}},\ }\href@noop {} {\emph {\bibinfo {title} {Capillarity and Wetting Phenomena}}}\ (\bibinfo  {publisher} {Springer New York, NY},\ \bibinfo {year} {2004})\ p.~\bibinfo {pages} {12}\BibitemShut {NoStop}%
\bibitem [{\citenamefont {Abramowitz}\ and\ \citenamefont {Stegun}(1948)}]{abramowitz1948handbook}%
  \BibitemOpen
  \bibfield  {author} {\bibinfo {author} {\bibfnamefont {M.}~\bibnamefont {Abramowitz}}\ and\ \bibinfo {author} {\bibfnamefont {I.~A.}\ \bibnamefont {Stegun}},\ }\href@noop {} {\emph {\bibinfo {title} {Handbook of mathematical functions with formulas, graphs, and mathematical tables}}},\ Vol.~\bibinfo {volume} {55}\ (\bibinfo  {publisher} {US Government printing office},\ \bibinfo {year} {1948})\BibitemShut {NoStop}%
\bibitem [{\citenamefont {Huh}\ and\ \citenamefont {Scriven}(1971)}]{huh-creep}%
  \BibitemOpen
  \bibfield  {author} {\bibinfo {author} {\bibfnamefont {C.}~\bibnamefont {Huh}}\ and\ \bibinfo {author} {\bibfnamefont {L.~E.}\ \bibnamefont {Scriven}},\ }\bibfield  {title} {\enquote {\bibinfo {title} {Hydrodynamic model of steady movement of a solid/liquid/fluid contact line},}\ }\href@noop {} {\bibfield  {journal} {\bibinfo  {journal} {J. Colloid Interface Sci.}\ }\textbf {\bibinfo {volume} {35}},\ \bibinfo {pages} {85--101} (\bibinfo {year} {1971})}\BibitemShut {NoStop}%
\bibitem [{\citenamefont {Voinov}(1977)}]{lmin}%
  \BibitemOpen
  \bibfield  {author} {\bibinfo {author} {\bibfnamefont {O.~V.}\ \bibnamefont {Voinov}},\ }\bibfield  {title} {\enquote {\bibinfo {title} {Hydrodynamics of wetting},}\ }\href@noop {} {\bibfield  {journal} {\bibinfo  {journal} {Fluid Dyn.}\ }\textbf {\bibinfo {volume} {11}},\ \bibinfo {pages} {714--721} (\bibinfo {year} {1977})}\BibitemShut {NoStop}%
\bibitem [{\citenamefont {Li}\ \emph {et~al.}(2023)\citenamefont {Li}, \citenamefont {Bodziony}, \citenamefont {Yin}, \citenamefont {Marschall}, \citenamefont {Berger},\ and\ \citenamefont {Butt}}]{friction}%
  \BibitemOpen
  \bibfield  {author} {\bibinfo {author} {\bibfnamefont {X.}~\bibnamefont {Li}}, \bibinfo {author} {\bibfnamefont {F.}~\bibnamefont {Bodziony}}, \bibinfo {author} {\bibfnamefont {M.}~\bibnamefont {Yin}}, \bibinfo {author} {\bibfnamefont {H.}~\bibnamefont {Marschall}}, \bibinfo {author} {\bibfnamefont {R.}~\bibnamefont {Berger}},\ and\ \bibinfo {author} {\bibfnamefont {H.-J.}\ \bibnamefont {Butt}},\ }\bibfield  {title} {\enquote {\bibinfo {title} {Kinetic drop friction},}\ }\href@noop {} {\bibfield  {journal} {\bibinfo  {journal} {Nat. Commun.}\ }\textbf {\bibinfo {volume} {14}},\ \bibinfo {pages} {4571} (\bibinfo {year} {2023})}\BibitemShut {NoStop}%
\bibitem [{\citenamefont {Quéré}(1999)}]{coating}%
  \BibitemOpen
  \bibfield  {author} {\bibinfo {author} {\bibfnamefont {D.}~\bibnamefont {Quéré}},\ }\bibfield  {title} {\enquote {\bibinfo {title} {Fluid coating on a fiber},}\ }\href@noop {} {\bibfield  {journal} {\bibinfo  {journal} {Annu. Rev. Fluid Mech.}\ }\textbf {\bibinfo {volume} {31}},\ \bibinfo {pages} {347--384} (\bibinfo {year} {1999})}\BibitemShut {NoStop}%
\bibitem [{\citenamefont {Gupta}\ \emph {et~al.}(2021)\citenamefont {Gupta}, \citenamefont {Konicek}, \citenamefont {King}, \citenamefont {Iqtidar}, \citenamefont {Yeganeh},\ and\ \citenamefont {Stone}}]{gravity-shape}%
  \BibitemOpen
  \bibfield  {author} {\bibinfo {author} {\bibfnamefont {A.}~\bibnamefont {Gupta}}, \bibinfo {author} {\bibfnamefont {A.~R.}\ \bibnamefont {Konicek}}, \bibinfo {author} {\bibfnamefont {M.~A.}\ \bibnamefont {King}}, \bibinfo {author} {\bibfnamefont {A.}~\bibnamefont {Iqtidar}}, \bibinfo {author} {\bibfnamefont {M.~S.}\ \bibnamefont {Yeganeh}},\ and\ \bibinfo {author} {\bibfnamefont {H.~A.}\ \bibnamefont {Stone}},\ }\bibfield  {title} {\enquote {\bibinfo {title} {Effect of gravity on the shape of a droplet on a fiber: Nearly axisymmetric profiles with experimental validation},}\ }\href@noop {} {\bibfield  {journal} {\bibinfo  {journal} {Phys. Rev. Fluids}\ }\textbf {\bibinfo {volume} {6}},\ \bibinfo {pages} {063602} (\bibinfo {year} {2021})}\BibitemShut {NoStop}%
\end{thebibliography}%

\end{document}